\documentstyle[aps,prl,twocolumn,epsfig]{revtex}
\input epsf         
\epsfverbosetrue    
\def\pss#1{\begin{center}\leavevmode \hbox{\epsfxsize=2in\epsfbox{#1}}\end{center}}
\def\ps#1{\begin{center}\leavevmode \hbox{\epsfxsize=2.5in\epsfbox{#1}}\end{center}}

\def\psbb#1{\begin{center}\leavevmode \hbox{\epsfxsize=6.6cm\epsfbox{#1}}\end{center}}
\def\fig#1{Fig.~\ref{#1}}
\def\refer#1{Ref.~\cite{#1}}
\def\eq#1{Eq.~(\ref{#1})}
\def\eqn#1{(\ref{#1})}
\def\bea {\begin{eqnarray}}
\def\eea {\end{eqnarray}}
\def\be {\begin{equation}}
\def\ee {\end{equation}}
\def\bk {\bf k}
\begin{document}
\draft
\title{Even-Odd and Super-Even Effects in the Attractive Hubbard Model}
\author{K. Tanaka and F. Marsiglio}
\address{Department of Physics, University of Alberta,
Edmonton, Alberta, Canada T6G 2J1}
\date{\today}
\maketitle
\begin{abstract}
The canonical BCS wave function is tested for the attractive Hubbard model.
Results are presented for one dimension, and are compared with the exact 
solutions by the Bethe ansatz and the results from the conventional grand 
canonical BCS approximation, for various chain lengths, electron densities, 
and coupling strengths. 
While the exact ground state energies are reproduced very well both by
the canonical and grand canonical BCS approximations, the canonical method
significantly improves the energy gaps for small systems and weak coupling.
The ``parity'' effect due to the number of electrons being even or odd 
naturally emerges in our canonical results.
Furthermore, we find a ``super-even'' effect:
the energy gap oscillates as a function of even electron number,
depending on whether the number of electrons is $4 m$ or $4 m + 2$ 
($m$ integer).  
Such oscillations as a function of electron number should be observable with 
tunneling measurements in ultrasmall metallic grains.
\end{abstract}

\makeatletter
\global\@specialpagefalse
\def\@oddhead{\hfill Alberta Thy 02-99}
\let\@evenhead\@oddhead
\makeatother

\section{INTRODUCTION}
\label{sec:int}

The possibility of fabricating \cite{ralph95} tunnel junctions containing
nanoscale particles has opened up new areas of exploration. Experiments
can now probe the energy spectrum in these particles and study changes as
a function of temperature and magnetic field \cite{black96,ralph97}. In this
way one can observe changes in the spectrum that are expected to occur
in the bulk due to phase transitions, and determine to what extent these
concepts are relevant for small systems \cite{anderson59}. The unprecedented
control over experimental conditions --- changes due to a single tunneling
electron can be observed --- allows one to ask and address questions which,
until now, have been academic only.\par

In this study we follow the experiments of Tinkham {\it et al.} 
\cite{ralph95,black96,ralph97} 
and address the issues concerning superconductivity in small systems.
Specifically they studied small
superconducting Al particles (diameter of order 10 nm) and probed, through
tunneling experiments, the excitation spectrum as a function of 
temperature and magnetic field. In attempting to treat small electronic
systems, there are many concerns related to possible surface and
impurity effects, both of which could lead to localization, for example.
For the present we ignore these potential complications, and instead
focus on a question which arises in the application of the
Bardeen-Cooper-Schrieffer (BCS) \cite{bardeen57} theory of superconductivity:
to what extent is the grand canonical ensemble useful (which in
the thermodynamic limit, is equivalent to the canonical one) 
for systems with a small number of electrons?

We propose to tackle this question in a systematic way, using the attractive
Hubbard model. The choice of this model is motivated by several factors. It
has long served as a paradigm for s-wave superconductivity, and serves as the
`minimal' model that best describes superconductivity. All the energy scales
are very well defined in the problem, so that no high frequency (and ill-defined)
cutoffs are required \cite{nozieres85}. Exact solutions are available in one
dimension via Bethe Ansatz techniques \cite{lieb68}, and the grand canonical
BCS solutions have recently been evaluated for large system sizes \cite{marsiglio97}.
In fact in that work some preliminary canonical solutions were also examined,
but only for very small system sizes. In this work we reformulate the canonical
solution, in such a way that larger systems up to three dimensions can be tackled.
We note that after this work was completed, a paper appeared \cite{braun98}
in which the canonical BCS equations were formulated and solved
for a model system with uniformly spaced level spacings. We will comment on
the similarities and differences with our own work in the course of our discussion
below.\par

The outline of the paper is as follows. In the next section we formulate the
problem. We mention the work of Falicov and Proetto \cite{falicov93}, who
applied a canonical BCS method to the problem of electrons on a tetrahedron (a
`small' fcc lattice). 
We have generalized this method for larger systems, and solved the ensuing equations numerically.
However, memory requirements become very demanding, and therefore, beyond a 
certain lattice size, we had to abandon this approach. 
We include it in the appendix,
nonetheless, because the variational wave function in this method
is more general than the BCS canonical wave function, and the variational
equation is linear.
In practice, however, as will be shown, this wave function only 
marginally improves 
the ground state energy.\par

We then formulate the canonical variational problem strictly in terms of the 
$g({\bf k})$'s familiar from the grand canonical solution \cite{schrieffer64}. 
The resulting equations are tedious, and can be cast in a more enlightening 
form by following the pioneering work of Dietrich, Mang and Pradal 
\cite{dietrich64}, who solved the canonical BCS problem for nuclei. 
This ``method of residues'' is based on representing expectation 
values in the BCS canonical ensemble through contour integrals of the BCS 
grand canonical ensemble expectation values:
the latter seems to have started with a paper by Bayman \cite{bayman60}.
This is also the methodology adopted by Braun and von Delft \cite{braun98},
who numerically evaluated residue integrals by fast Fourier transform.
As will be explained, however, we chose to evaluate the required integrals
analytically (which requires numerical summations).
A concise summary of the residue method and further references
are available in Ref. \cite{ringschuck}. \par

Finally, we provide a review of the grand canonical formulation and the
Bethe Ansatz solutions in one dimension \cite{lieb68,bahder86,marsiglio97},
for purposes of comparison. We have also performed some numerical diagonalizations,
although these are limited to small systems and (in one dimension)
serve only to verify the Bethe Ansatz solutions. In both the grand canonical
and canonical BCS 
formulations we have performed the variation strictly in terms of the
single set of parameters,  $g({\bf k})$. This departs from more conventional
treatments where two sets of parameters, $u({\bf k})$, and $v({\bf k})$, are used,
with an auxiliary relation between them.  Our formulation is conceptually clearer,
but slightly more cumbersome, so some details are provided. \par

The following section is devoted to numerical results. All the 
computational results presented in this paper
will be in one dimension. This is because the exact results for `large' systems
are only available in 1-D, and therefore it is presently only in this case that one can
study the full crossover from the bulk limit to small systems. Because the attractive
Hubbard model is a very local model, many of the features of the variational
solution will remain in higher dimension; in fact we expect the agreement
of the canonical BCS results with the exact ones to improve, but we have
no way at present to check this. In addition, many of the features
of the solutions in higher dimensions (such as long range order) will not be
present in one dimension. A more systematic study of the canonical BCS solution
in two and three dimensions will be presented elsewhere (unfortunately without
exact checks, except for very small systems).\par

Aside from evaluations of the canonical BCS formulation we will demonstrate
the so-called `parity effect' \cite{tuominen92,lafarge93}, where qualitative
differences in the tunneling gap can be observed, depending on whether an
even or odd number of electrons occupies the small Al particle. While this
has been understood within a parity-conserved grand canonical BCS formulation
\cite{janko94}, it becomes much clearer in a canonical formulation. In addition
we have found an interesting $4 m$ vs. $4 m + 2$ effect, 
where $m$ is an integer. 
In our case we have observed oscillations as a function of electron number.
Therefore we have variations in the gap between odd, even and {\it 
super-even} (i.e. multiples of 4) electrons. Hints of
such behaviour, noted as a function of {\it lattice size}, were first
discussed by Fye {\it et al.} \cite{fye90}.\par

The concluding section summarizes our results.  In the Appendix we outline
the formulation of the linearized canonical variation.
\par

\section{FORMULATION}
\label{sec:form}
\subsection{Model}
\label{model}

The attractive Hubbard Hamiltonian is given by
\begin{mathletters}
\begin{eqnarray}
H&=&-\,\sum_{i,\delta \atop \sigma}t_\delta\,(a_{i+\delta,\sigma}^\dagger a_{i\sigma} 
+{\rm h.c.}) - |U| \sum_{i} n_{i\uparrow} n_{i\downarrow}
\label{hamilc}\\
&=&\sum_{\bf{k}\sigma} \epsilon_{\bf{k}}\,a_{\bf{k}\sigma}^\dagger
a_{\bf{k}\sigma} - {|U|\over N} \sum_{\bf{k}\bf{k}'\bf{q}}
\,a_{{\bf k}\uparrow}^\dagger
a_{-{\bf k} + {\bf q}\downarrow}^\dagger a_{-{\bf k}' + {\bf q} \downarrow} 
a_{{\bf k}'\uparrow}
\;,\label{hamilk}
\end{eqnarray}
\end{mathletters}
where $a_{i\sigma}^\dagger$ ($a_{i\sigma}$) creates (annihilates) an electron 
with spin $\sigma$ at site $i$
and $n_{i\sigma}$ is the number operator for an electron with spin $\sigma$ 
at site $i$ ($i$ is the index for the primitive vector ${\bf R}_i$).
The $t_\delta$ is the hopping rate of electrons from one site to a neighbouring
site ${\bf R}_\delta$ away (often nearest neighbours only are included),
and $|U|$ is the coupling strength between electrons on the same site;
here the fact that it is attractive is explicitly included from the beginning.
In \eq{hamilk}, 
we have Fourier transformed the Hamiltonian with periodic boundary conditions
for $N$ sites in each dimension.
The $a_{{\bf k}\sigma}^\dagger$ and $a_{{\bf k}\sigma}$ are the creation and 
annihilation operators in the reciprocal space, 
and the kinetic energy is given by
\begin{equation}
\epsilon_{\bf{k}}=-2\,\sum_{\delta}t_\delta\,{\rm cos}\,{\bf k}\cdot{\bf R}_\delta\;,
\label{kenergy}
\end{equation}
where ${\bf R}_\delta$ is the coordinate vector connecting sites $i$ to $i+\delta$.

We perform variational calculations using the component of the
BCS wave function that has a given number of pairs $\nu$
and thus conserves the number of electrons $N_e=2\nu$.
The BCS wave function is a superposition of pair states with all the possible
numbers of pairs $\{\nu\}$:
\begin{equation}
|{\rm BCS}\rangle_{\rm GC} =\prod_{\bf{k}}\,\bigl(\,u_{\bf k}+v_{\bf k}\,
a_{{\bf k}\uparrow}^\dagger a_{-{\bf k}\downarrow}^\dagger\,\bigr)\,|0\rangle
\equiv \sum_{\nu=0}^{\infty}|\Psi_{2\nu}\rangle\;,
\label{bcswf}
\end{equation}
where $|0\rangle$ denotes the vacuum state, and 
$|\Psi_0\rangle\equiv |0\rangle$.
The $\nu$-pair component $|\Psi_{2\nu}\rangle$ can be obtained by 
rearranging $|{\rm BCS}\rangle_{\rm GC}$ into a power series of the pair creation
operator $a_{{\bf k}\uparrow}^\dagger a_{-{\bf k}\downarrow}^\dagger$
and can be written as 
\begin{eqnarray}
|\Psi_{2\nu}\rangle&=&
\sum_{{\bf k}_1\,<}\sum_{{\bf k}_2\,<}\cdots\sum_{<\,{\bf k}_\nu}\,
\prod_{i=1}^{\nu}\,\Bigl(\,g({{\bf k}_i})\,a_{{\bf k}_i\uparrow}^\dagger 
a_{-{\bf k}_i\downarrow}^\dagger\,\Biglb)\,|0\rangle\;\label{cbcswf}\\
&=&{1\over \nu !}\,\prod_{i=1}^{\nu}\,
\Bigl(\,\sum_{{\bf k}_i} g({{\bf k}_i})\,a_{{\bf k}_i\uparrow}^\dagger a_{-{\bf k}_i\downarrow}^\dagger
\,\Biglb)\,|0\rangle\nonumber\\
&&({\bf k}_i\neq {\bf k}_j\,,\;{\rm for\,all}\; i\not= j)\;.\eqnum{4$^\prime$}
\label{cbcswfp}
\end{eqnarray}
Here we have defined 
$g({{\bf k}_i})=\Bigl(\prod_{\bf k} u_{\bf k}\Biglb)^{1/\nu}v_{{\bf k}_i}/u_{{\bf k}_i}$.
The condition ${\bf k}_1<{\bf k}_2<\cdots<{\bf k}_\nu$ 
for the $\{{\bf k}_i\}$ sums in \eq{cbcswf} means that 
all the ${\bf k}_i$'s should be different and ordered so that
all the different combinations of $\{{\bf k}_i\}$ are counted once each:
the latter condition is removed for the sums in \eq{cbcswfp} and 
the division by $\nu !$ compensates for the multiple counting.
In a finite system with $N$ sites in each dimension, the wave number can
take $N^D$ values, where $D$ is the dimension of the system, and 
hence the $\nu$ sums over $\bf k$'s in \eq{cbcswf} result in 
$_{N^D}C_\nu={N^D ! \over \nu !\,(N^D-\nu) !}$ terms.

\subsection{Grand Canonical Variation}
\label{bcs}

We begin by reviewing the grand canonical formulation of BCS theory. We have opted,
contrary to convention, to formulate the minimization problem in
terms of only the variational parameters, $g_{\bk} \equiv g(\bk)$ (and not the $u_{\bk}$'s and $v_{\bk}$'s).
This was originally motivated by the loss of clear meaning for the $u_{\bk}$'s and $v_{\bk}$'s
in the canonical ensemble, though in fact one can proceed equally well with these
\cite{dietrich64,braun98}. One therefore simply writes:
\be
|{\rm BCS}\rangle_{\rm GC} = c \prod_{\bf k}\,\bigl(\,1+g_{\bk} \,
a_{{\bf k}\uparrow}^\dagger a_{-{\bf k}\downarrow}^\dagger\,\bigr)\,|0\rangle
\;,\label{bcswfg}
\ee
where the coefficient $c$ properly normalizes the wave function. One then calculates
the various expectation values to determine the ground state energy:
\begin{equation}
E = {\langle\Psi|H|\Psi\rangle\over
\langle\Psi|\Psi\rangle}\;,
\label{energy}
\end{equation}
where, in this case, the wave function is the BCS grand canonical one given in
Eq. (\ref{bcswfg}). For example, the expectation value of the kinetic energy 
operator, $\hat{T}$ is
\begin{equation}
\langle\Psi|\hat{T}|\Psi\rangle = 2\,|c|^2\,\prod_{\bf p} (1 + g_{\bf p}^2)
\sum_{\bk}\,(\epsilon_{\bk} - \mu)\,{g_{\bk}^2 \over 1 + g_{\bk}^2}\;.
\label{kinetic}
\end{equation}
Note that normally the product and the denominator are absent because these
factors are normally defined to be unity due to normalization. Also note that
the $g_{\bk}$'s are generally complex, but can be chosen to be real; this is
presumed in Eq. (\ref{kinetic}), and will be implicit in what follows. Careful
evaluation of the potential energy terms yields the following result:
\bea
&E_{\rm GC}&=2\,\sum_{\bf k}\,(\epsilon_{\bf k} - \mu)\,{g_{\bk}^2\over 1+ g_{\bk}^2} - {|U|\over N}\,\sum_{\bk} {g_{\bk}^2\over 1+g_{\bk}^2}\nonumber\\
&-&{|U|\over N}\,\sum_{\bk \not= {\bf k}^\prime}{g_{\bk}  \over 1+g_{\bk}^2} {g_{{\bf k}^\prime}  \over 1+g_{{\bf k}^\prime}^2} 
- {|U|\over N}\,\sum_{\bk \not= {\bf k}^\prime} {g_{\bk}^2\over 1+g_{\bk}^2} {g_{{\bf k}^\prime}^2\over 1+g_{\bf k^\prime}^2}
\label{grand_e}
\eea
with $\mu$ the
chemical potential; it plays the mathematical role of the Lagrange multiplier for the condition
that the average electron number $_{\rm GC}\langle{\rm BCS}|\hat{N}_e|{\rm BCS}\rangle_{\rm GC} = N_e$.
We have written the total energy term by term in the following sequence: the first term is the
kinetic energy, the following two come from Cooper pair scattering (${\bf q} = 0$), and the
fourth term is the Hartree term (${\bf q} \not= 0$). When displayed this way it is apparent
that the last term excludes the spurious Hartree term whereby an electron interacts with
itself (a `$1/N$ effect'). It is also apparent that the {\it reduced} BCS Hamiltonian (${\bf q} = 0$ 
scattering only) will exclude this last term, and in doing so omits the Hartree term (since it 
is often deemed to merely represent a shift in energies). At the same time Eq. (\ref{grand_e})
appears to have terms of order $1/N$ smaller than the dominant terms. In fact it does not, 
and we can rewrite this equation in the following form: 
\bea
&E_{\rm GC}&= 2 \, \sum_{\bf k}\,(\epsilon_{\bf k} - \mu)\,{g_{\bk}^2\over 1+ g_{\bk}^2}\nonumber\\
&-&{|U|\over N}\,\sum_{\bk, {\bf k}^\prime} \biggl[\,{g_{\bk} \over 1+g_{\bk}^2} 
{g_{{\bf k}^\prime} \over 1+g_{{\bf k}^\prime}^2}
+{g_{\bk}^2\over 1+g_{\bk}^2} {g_{{\bf k}^\prime}^2\over 1+g_{{\bf k}^\prime}^2}\,\biggr]\;,
\label{grand_e2}
\eea
where now the summations are unrestricted. The entire expression is now clearly extensive,
as it should be, though in this form {\it pairing} terms of order $1/N$ have been mixed in 
with the Hartree term.

The next step is to carry out the variation with respect to the $g_{\bk}$'s; this can be
done straightforwardly to yield:
\be
2\,(\epsilon_{\bk} - \tilde\mu)\,g_{\bk} =  
{|U|\over N}\,\sum_{{\bf k}^\prime}\,
{g_{{\bf k}^\prime} \over 1+g_{{\bf k}^\prime}^2}
\,\bigl[ 1 - g_{\bk}^2 \bigr]\;,
\label{bcs_gc}
\ee
where
\be
\tilde\mu \equiv \mu + {|U|\over N}\,\sum_{\bk} {g_{\bk}^2\over 1+ g_{\bk}^2}\;.
\label{hart}
\ee
We also define the pair potential,
\be
\Delta_{\rm BCS} \equiv {|U|\over N}\,\sum_{\bk} {g_{\bk} \over 1+ g_{\bk}^2}  
\label{pairpot}
\ee
so that the BCS equation can be written
\be
2\,(\epsilon_{\bk} - \tilde\mu)\,g_{\bk} = \Delta_{\rm BCS}\,\bigl[ 1 - g_{\bk}^2 \bigr]\;.
\label{bcs_gc1}
\ee
The solution is 
\be
g_{\bk} = {E_{\bk} - (\epsilon_{\bk} - \tilde\mu) \over \Delta_{\rm BCS}}\;,
\label{soln}
\ee
where $E_{\bk} \equiv \sqrt{ (\epsilon_{\bk} - \tilde\mu)^2 + \Delta_{\rm BCS}^2}$ is the quasi-particle
energy. This solution is only implicit, since $\Delta_{\rm BCS}$ depends on the $g_{\bk}$'s
through Eq. (\ref{pairpot}), and must be determined by numerical iteration. The number equation
is also required to determine the chemical potential 
as a function of coupling strength. It is \cite{schrieffer64}
\be
n \equiv {\langle N_e\rangle \over N} = 1 - {1 \over N} \sum_{\bk}\,{(\epsilon_{\bk} - \tilde\mu) \over E_{\bk}}\;.
\label{bcs_num}
\ee
The gap is then defined to be
\be
\Delta_{\circ} = {\rm min}\,(E_{\bk})\;.
\label{mingap}
\ee
If we define the minimum band energy, $\epsilon_{\rm min} \equiv -D/2$, 
with $D$ the bandwidth, then it follows that $\Delta_{\circ}=\sqrt{(\epsilon_{\rm min} - \tilde\mu)^2 + \Delta_{\rm BCS}^2}$ when $\tilde\mu < \epsilon_{\rm min}$.
This definition implies that 
the gap depends on a certain momentum, determined by the wave vector $\bf k$
at which the minimum energy occurs.
The total energy also follows readily
from these expressions. It is
\be
{E_{\rm GC} \over N} = {1 \over N} \sum_{\bk} \epsilon_{\bk} \biggl(
1 - {\epsilon_{\bk} - \tilde\mu \over E_{\bk}} \biggr) - |U| \biggl({n \over 2}
\biggr)^2 - { \Delta_{\rm BCS}^2 \over |U|},
\label{bcs_gc2}
\ee
where we have used the fact that $n = {2 \over N} \,\sum_{\bk} {g_{\bk}^2 \over 1+ g_{\bk}^2}$.
The BCS grand canonical results in the remainder of the paper have been obtained through the
solution of these equations on finite lattices. 

\subsection{Canonical Variation}
\label{canform}

\subsubsection{BCS}
\label{bcscanform}

Rather than adopt the wave function given by Eq. (\ref{bcswfg}), which spans all (even) number
sectors, one can work directly with the wave function $|\Psi_{2\nu}\rangle$ given by Eq.
(\ref{cbcswf}) with fixed electron number, $N_e = 2\nu$. A `simplification' occurs if
we try to linearize the problem, by defining a variational parameter 
$C({\bf k}_1,{\bf k}_2,\cdots,{\bf k}_\nu)\equiv g({\bf k}_1)\,g({\bf k}_2)
\cdots g({\bf k}_\nu)$ for this wave function, following Refs.\cite{falicov93} 
and \cite{marsiglio97} for the two-pair case.
Thus one can rewrite \eq{cbcswf} as
\be
|\Psi_{2\nu}\rangle=
\sum_{{\bf k}_1\,<}\sum_{{\bf k}_2\,<}\cdots\sum_{<\,{\bf k}_\nu}\,
C({\bf k}_1,{\bf k}_2,\cdots,{\bf k}_\nu)
\prod_{i=1}^{\nu}\,a_{{\bf k}_i\uparrow}^\dagger
a_{-{\bf k}_i\downarrow}^\dagger\,|0\rangle\;.\label{cwfe}
\ee
As discussed in \refer{falicov93}, the
$C$'s defined as above are subject to constraints; for example,
for $\nu=2$ and for a given set of four ${\bf k}$ values
(${\bf q}_1,{\bf q}_2,{\bf q}_3,{\bf q}_4$) which satisfy
${\bf q}_1<{\bf q}_2<{\bf q}_3<{\bf q}_4$,
\bea
C({\bf q}_1,{\bf q}_2)\,C({\bf q}_3,{\bf q}_4)&=&
C({\bf q}_1,{\bf q}_3)\,C({\bf q}_2,{\bf q}_4)\nonumber\\
&=&C({\bf q}_1,{\bf q}_4)\,C({\bf q}_2,{\bf q}_3)\;.
\eea
If these constraints are ignored and the
$C$'s in \eq{cwfe} are treated as $_{N^D}C_\nu$ independent parameters,
the minimization of the energy 
\begin{equation}
E_{2\nu}={\langle\Psi_{2\nu}|H|\Psi_{2\nu}\rangle\over 
\langle\Psi_{2\nu}|\Psi_{2\nu}\rangle}
\label{venergy}
\end{equation}
with respect to $\{C\}$ reduces to a linear problem \cite{falicov93}.
We initially adopted this method of canonical variation, which actually
allows more variational freedom than using the $g_{\bk}$'s. However, as
one can readily appreciate, the number of variational parameters grows
very quickly with increasing lattice size, and therefore in practice, 
this method
has limited usage. Where it was practical, however, we carried out these
calculations and compared them with the canonical BCS results to be
described below. The gain in accuracy for the ground state energy
was fairly small. Our formulation and results in this case are summarized in 
the Appendix.

The cost of working with the $g_{\bk}$'s is that the problem remains a very
nonlinear one. Nonetheless the gain in computational ease more than compensates 
for this, in that
only $N$ variational parameters are required, where $N$ is the number of lattice sites.
As we shall see, this allows us to easily study the limit from small systems
to those that are well described by the grand canonical ensemble.
The formulation of this problem using the expression for $|\Psi_{2\nu}\rangle$
given in \eq{cbcswf} is straightforward but tedious.
We therefore follow the ``method of residues''
\cite{dietrich64} (see also \cite{ringschuck} and
more recently \cite{braun98}) originally developed for the nuclear pairing problem.
An advantage of this method is that matrix elements are more easily evaluated (they
remain as simple as the ones in the grand canonical formulation). Moreover 
the energy and the resulting variational equation 
can be written in a compact way which parallels the grand canonical BCS equations.
The starting point is to write \eq{cbcswf} in the general form as a
particle-number projection of the grand canonical BCS state \cite{bayman60}.
In more general terms, this is the projection that restores the symmetry of the Hamiltonian
(i.e., conserved particle number in this case) in the wave function \cite{ringschuck,braun98a}.

For an even number of electrons the
$\nu$-pair wave function in the projected form is
\begin{equation}
|\Psi_{2\nu}\rangle={1\over 2\pi i}\oint d\xi\;\xi^{-\nu-1}
\prod_{\bf k}\,\Bigl(\,1+\xi\,g_{\bf k}\,a_{{\bf k}\uparrow}^\dagger 
a_{-{\bf k}\downarrow}^\dagger\,\Bigr)\,|0\rangle\;,
\label{pbcswfe}
\end{equation}
where the contour is any counterclockwise path that encloses the origin.
If we perform the residue integral above, 
we recover the desired wave function \eq{cbcswf}.

As written above the wave function is not normalized, so expectation values of observables will include
a factor given by $\langle\Psi_\nu|\Psi_\nu\rangle$ in the denominator.
A straightforward evaluation of this factor gives $\langle\Psi_\nu|\Psi_\nu\rangle = R_0^0$,
where $R_n^m$ is defined by the residue integral \cite{dietrich64}
\bea
&&R_n^m({\bf k}_1,{\bf k}_2,\cdots,{\bf k}_m)\nonumber\\
\nonumber\\
&=&{1\over 2\pi i}\oint d\xi\;\xi^{-(\nu-n)-1}
\prod_{{\bf k}\neq{\bf k}_1,{\bf k}_2,\cdots,{\bf k}_m}
\bigl(\,1+\xi\,g_{\bf k}^2\,\bigr)\label{residuea}\\
\nonumber\\
&=&\sum_{{\bf p}_1\,<}\sum_{{\bf p}_2\,<}\cdots\sum_{<\,{\bf p}_{\nu-n}}\,
g_{{\bf p}_1}^2 g_{{\bf p}_2}^2 \cdots g_{{\bf p}_{\nu-n}}^2\;,
\label{residueb}\\
\nonumber\\
{\rm where}&&{\bf p}_i\neq {\bf k}_1,{\bf k}_2,\cdots,{\bf k}_m\quad(i=1,\cdots,\nu-n)\;.
\nonumber
\eea
The integral in \eq{residuea} has sharp oscillations as a function of 
$\xi$ (or $\theta$ defined by the coordinate transformation $\xi=e^{i\theta}$)
whose severity increases with increasing system size.
We therefore evaluated the integral using the analytical result \eq{residueb}.
This was accomplished efficiently with an algorithm that took advantage of
the polynomial structure of the original integrand in \eqn{residuea}, and led to a considerable
speed up in the integral evaluations. 

These residue integrals are useful not only because the matrix elements can
be easily evaluated, but also because they satisfy various recursion and
sum formulas. For example, Dietrich {\it et al.} \cite{dietrich64} found the
following recursion relation 
\bea
R_n^m({\bf k}_1,{\bf k}_2,\cdots,{\bf k}_m)&=&R_n^{m+1}({\bf k}_1,{\bf k}_2,
\cdots,{\bf k}_m,{\bf k})\nonumber\\&+&g_{\bf k}^2\,R_{n+1}^{m+1}({\bf k}_1,
{\bf k}_2,\cdots,{\bf k}_m,{\bf k})\;.
\label{recursion}
\eea
This result follows straightforwardly from the definition. Another useful
relation involves the derivative required in the variational principle:
\be
{\partial R_n^m({\bf k}_1,{\bf k}_2,\cdots,{\bf k}_m)\over \partial g_{\bf k}}
=2\,g_{\bf k}\,R_{n+1}^{m+1}({\bf k}_1,{\bf k}_2,\cdots,{\bf k}_m,{\bf k})\;.
\label{derivative}
\ee
In addition, we have also found and exploited the following sum rule:
\bea
\sum_{{\bf k}\neq{\bf k}_1,{\bf k}_2,\cdots,{\bf k}_m}
g_{\bf k}^2\,&&R_{n+1}^{m+1}({\bf k}_1,{\bf k}_2,\cdots,{\bf k}_m,{\bf k})
\nonumber\\
=(\nu-n)\,&&R_n^m({\bf k}_1,{\bf k}_2,\cdots,{\bf k}_m)\quad(\nu>n)\;.
\label{sumrule}
\eea
All these relations follow in a very straightforward fashion from the
definition.  For completeness we include the case where the residue 
with two or more equal indices is required; it is then useful to rewrite
the definition of the residue, \eq{residuea} as
\bea
&&R_n^m({\bf k}_1,{\bf k}_2,\cdots,{\bf k}_m)\nonumber\\\nonumber\\
&=&{1\over 2\pi i}\oint d\xi\;\xi^{-(\nu-n)-1}
{\prod_{\bf k}\bigl(\,1+\xi\,g_{\bf k}^2\,\bigr)\over 
\prod_{{\bf k}={\bf k}_1,{\bf k}_2,\cdots,{\bf k}_m}
\bigl(\,1+\xi\,g_{\bf k}^2\,\bigr)}\label{residuee}\;.
\eea
The advantage of this definition over \eq{residuea} is that there is
no ambiguity when one or more momentum index is used more than once.
With this definition
these residues still satisfy the recursion relation \eqn{recursion}.
This fact is useful for simplifying the 
expressions that involve restricted double momentum sums, which
arise in the canonical formulation.

Omitting details, the ground state energy can be written in a
form reminiscent of the grand canonical BCS
formulation (see \eq{grand_e2}):
\bea
E_{2\nu}&=&\sum_{\bf k}\,2\,\epsilon_{\bf k}\,{g_{\bf k}^2 \over 1 + g_{\bf k}^2}\,r_1^1({\bf k})\nonumber\\
&-&{|U|\over N}\sum_{{\bf k}, {\bf k}^\prime}\,\biggl[\,
{g_{\bf k} \over 1 + g_{\bf k}^2}\,{g_{{\bf k}^\prime} \over 1 + g_{{\bf k}^\prime}^2}\,r_1^2({\bf k},{{\bf k}^\prime})\nonumber\\ 
&&\quad\quad\quad\,+\,
{g_{\bf k}^2 \over 1 + g_{\bf k}^2}\, {g_{{\bf k}^\prime}^2 \over 1 + g_{{\bf k}^\prime}^2}
\,r_2^2({\bf k},{\bf k}^\prime)\,\biggr]\;,
\label{gform_e}
\eea
where we have now defined normalized residues:
\be
r_n^m({\bf k}_1, \cdots, {\bf k}_m) \equiv { R_n^m({\bf k}_1, \cdots, {\bf k}_m) \over R_0^0}
\prod_{{{\bf k}^\prime}={\bf k}_1,\cdots,{\bf k}_m} \bigl( 1 + g_{{\bf k}^\prime}^2 \bigr).
\label{normres}
\ee 
This definition is clearly motivated by the fact that in the bulk limit, the
canonical results converge to the grand canonical ones.

A straightforward but tedious variation of \eq{gform_e} yields the following variational 
equation:
\be
(\,2\tilde\epsilon_{\bf k}+\Lambda_{\bf k}\,)g_{\bk} = \Delta_{\bf k}\,
\bigl[\,1 - g_{\bf k}^2\,\bigr]\;,
\label{gsol}
\ee
where
\bea
\tilde\epsilon_{\bf k}& \equiv &\,\epsilon_{\bf k}\,r_1^1({\bk}) -
{|U|\over N}\, \sum_{{\bf k}^\prime} {g_{{\bf k}^\prime}^2 \over 1 + g_{{\bf k}^\prime}^2}\,r_2^2({\bf k}^\prime,{\bk})
\label{etilde_ce}\\
\nonumber\\
\Delta_{\bf k}& \equiv &{|U|\over N}\sum_{{\bf k}^\prime}\,{g_{{\bf k}^\prime} 
\over 1 + g_{{\bf k}^\prime}^2}\, r_1^2({\bf k}^\prime,{\bf k})
\label{delta_ce}
\eea
\bea
\Lambda_{\bf k}& \equiv & 2\,\epsilon_{\bf k}\,g^2_{\bk}\,r_1^1({\bk})\,\bigl[ 1 - r_1^1({\bk}) \bigr]\nonumber\\ 
&+&{|U| \over N} \biggl\{ r_1^1({\bk}) ( 1 -  g^2_{\bk} ) - r_1^2({\bk},{\bk}) \biggr\}\nonumber\\
&+& (1 + g^2_{\bk} ) \sum_{{\bf k}^\prime \neq {\bk}}\,2\,\epsilon_{{\bf k}^\prime}\,{g^2_{{\bf k}^\prime} \over 1 + g^2_{{\bf k}^\prime}}
\biggl( r_2^2({\bf k}^\prime, {\bk}) - r_1^1({\bk}) r_1^1({\bf k}^\prime) \biggr)\nonumber\\
&-&2\,{|U| \over N}\,g_{\bk} \sum_{{\bf k}^\prime \neq {\bk}} {g_{{\bf k}^\prime} \over 1 + g^2_{{\bf k}^\prime}} 
\biggl( r_1^2({\bf k}^\prime, {\bk}) + g_{{\bf k}^\prime} g_{\bk} r_2^2({\bf k}^\prime,{\bk}) \biggr)\nonumber\\
&-& {|U| \over N} (1 + g^2_{\bk} ) \sum_{{\bf p},{\bf p}^\prime \neq {\bk}} 
{g_{\bf p} \over 1 + g^2_{\bf p}} {g_{{\bf p}^\prime} \over 1 + g^2_{{\bf p}^\prime}}
\biggl[\,r_2^3({\bf p},{\bf p}^\prime,{\bk})\nonumber\\
&&\hspace{3cm}+\,g_{\bf p}\,g_{{\bf p}^\prime}\,r_3^3({\bf p},{\bf p}^\prime,{\bk})\,\biggr]\nonumber\\
&+&{|U| \over N} (1 + g^2_{\bk})\,r_1^1({\bk})\, \sum_{{\bf p}, {\bf p}^\prime}
{g_{\bf p} \over 1 + g^2_{\bf p}} {g_{{\bf p}^\prime} \over 1 + g^2_{{\bf p}^\prime}}\biggl[\,r_1^2({\bf p},{\bf p}^\prime)\nonumber\\
&&\hspace{3cm}+\,g_{\bf p}\,g_{{\bf p}^\prime}\,r_2^2({\bf p},{\bf p}^\prime)
\,\biggr] \label{lambda_ce}
\eea
A solution to Eq. (\ref{gsol}) is obtained by numerical iteration. Note that \eq{gsol}
resembles the analogous equation in the grand canonical ensemble, \eq{bcs_gc1}. The factor
$\Lambda_{\bk}$ appears here in addition; all terms in \eq{lambda_ce} are of order $1/N$,
and therefore vanish in the thermodynamic limit. Moreover, both the single-particle energy
\eq{etilde_ce} and the pairing potential \eq{delta_ce} are modified by the normalized
residue integrals.
In practice, instead of Eqs.~(\ref{gsol})-(\ref{lambda_ce}), we used the simpler-looking expression:
\bea
g_{\bf k}&=&-{|U| \over N}\,{\sum_{{\bf p}\neq{\bk}}R_1^2({\bk},{\bf p})
\over {\rm denom}}\;,\quad{\rm where}\nonumber\\\nonumber\\
{\rm denom}&=&\biggl(E_{2\nu}-2\epsilon_{\bk}+{|U|\over N}\biggr)\,R_1^1({\bk})
\nonumber\\
&-&\sum_{{\bf p}\neq{\bk}}\biggl(2\epsilon_{\bf p}-3{|U|\over N}\biggr)\,g_{\bf p}^2\,R_2^2({\bk},{\bf p})\nonumber\\
&+&{|U|\over N}\sum_{{\bf p}\neq{\bk}}\sum_{{\bf p}^\prime\neq{\bk},\,{\bf p}^\prime\neq{\bf p}} \biggl[\,g_{\bf p}\,g_{{\bf p}^\prime}\,R_2^3({\bk},{\bf p},{\bf p}^\prime)\nonumber\\&&\hspace{2.6cm}+\,g_{\bf p}^2\,g_{{\bf p}^\prime}^2\,R_3^3({\bk},{\bf p},{\bf p}^\prime)\,\biggr]\;.
\eea
Although this equation does not resemble the grand canonical equations,
it is homogeneous in $\{g_{\bf k}\}$  and has far fewer terms.

In the case of an odd electron number $N_e=2\nu+1$, we define the fixed $N_e$ wave function
in terms of a residue integral as
\be
|\Psi_{2\nu+1}\rangle={1\over 2\pi i}\oint d\xi\;\xi^{-\nu-1}\,
a_{{\bf q}\sigma}^\dagger\prod_{{\bf k}\neq{\bf q}}\,
\Bigl(\,1+\xi\,g_{\bf k}\,a_{{\bf k}\uparrow}^\dagger 
a_{-{\bf k}\downarrow}^\dagger\,\Bigr)\,|0\rangle\;.
\label{pbcswfo}
\ee
This wave function carries a momentum label {\bf q} which gives the momentum
of the unpaired electron (and hence of the total state).
The normalization factor is now $\langle\Psi_{2\nu+1}|\Psi_{2\nu+1}\rangle
=R_0^1({\bf q})$, and the total energy is given by
\bea
E_{2\nu+1}& = &\epsilon_{\bf q}+\sum_{{\bf k}\neq{\bf q}}
\biggl(\,2\,\epsilon_{\bf k}-{|U|\over N}\,\biggr)\,
{g_{\bf k}^2 \over 1 + g_{\bf k}^2}\,{r_1^2({\bf k},{\bf q}) \over r_0^1({\bf q})}\nonumber\\\nonumber\\
& - &{|U|\over N}\sum_{{\bf k}\neq{\bf q}}\sum_{{\bf k}^\prime\neq{\bf q}}
\biggl[\,{g_{\bf k} \over  1 + g_{\bf k}^2}\,{g_{{\bf k}^\prime} \over  1 
+ g_{{\bf k}^\prime}^2}\,{r_1^3({\bf k},{\bf k}^\prime,{\bf q})
\over r_0^1({\bf q})}\nonumber\\\nonumber\\
&&\hspace{1.8cm}+\,
{g_{\bf k}^2 \over  1 + g_{\bf k}^2} \, {g_{{\bf k}^\prime}^2 \over  1 
+ g_{{\bf k}^\prime}^2}
\,{r_2^3({\bf k},{\bf k}^\prime,{\bf q})  \over r_0^1({\bf q})}\,
\biggr]\,.
\label{gform_eo}
\eea
This expression for the ground state energy has the same form as \eq{gform_e}
except that the kinetic energy of the unpaired electron is singled out and the
momentum sums explicitly prohibit the singly occupied momentum state. As in the even case
a lengthy variational equation for the $g_{\bk}$'s is obtained and must be
solved numerically. The end result is that we have the ground state energy for any number
of electrons. 

Thus one can construct the gap by
\be
2\Delta(N_e) \equiv E_{N_e - 1} - 2E_{N_e} +  E_{N_e + 1}.  
\label{cangap}
\ee
Various definitions of a gap or binding energy exist in the literature 
\cite{hirsch88,fye90,marsiglio97,matveev97}. The basic idea is the same --- one
wants to compare the difference in energies between two systems, one in which
$2N_e$ electrons are distributed equally over two subsystems containing $N_e$ electrons
each, and the other in which the subsystems contain $N_e + 1$ and $N_e -1$ electrons
respectively. If $N_e$ is even, the former has lower energy since
pairing is fully utilized; this is reflected in a positive gap. If $N_e$ is
odd the latter has lower energy and therefore the gap is negative. This is the origin of
the so-called parity effect measured by Tinkham and coworkers \cite{ralph95,black96,ralph97,tuominen92}.
For evaluating the gap $\Delta(N_e)$ in \eq{cangap},  
we choose the momentum $\bf q$ that yields the lowest energy for a given
odd number of electrons. This is important for properly recovering the exact results
in certain regimes, i.e. strong coupling and/or low electron density.

In the grand canonical BCS formulation, there is a direct correspondence
between the variational parameters and the occupation probability 
$n_{\bk} \equiv \sum_\sigma 
\langle a_{{\bf k} \sigma}^\dagger a_{{\bf k} \sigma} \rangle$.
It is determined through the $g_{\bf k}$'s as
\be
n_{\bk} = 2\,{g_{\bk}^2 \over 1 + g_{\bk}^2} = 1 - {\epsilon_{\bk} - \tilde\mu
\over E_{\bk}}\;,
\label{occ_gc}
\ee  
where the various functions are defined in Section \ref{bcs}. Utilizing the $u_{\bk}$'s
and $v_{\bk}$'s instead of the $g_{\bk}$'s makes the correspondence 
even more transparent, for we have
$n_{\bk} = 2v_{\bk}^2$ in that case. In the canonical formulation,
for an even number of electrons, for example, we construct the matrix element
\be
n_{{\bk}\sigma} = \langle \psi_{2\nu} |\,a_{{\bf k} \sigma}^\dagger a_{{\bf k} \sigma}\,| \psi_{2\nu} \rangle,
\label{occ_c}
\ee
and obtain
\be
n_{\bk} = \sum_{\sigma} n_{{\bf k}\sigma}= 2\,{g_{\bk}^2 \over 1 + g_{\bk}^2}
\,r_1^1({\bk})\,.
\label{occ_ca}
\ee

\subsubsection{{Exact Solutions}}
\label{exactcanform}

The formalism developed so far is applicable in any dimension. The results to be discussed
later in this paper focus on one dimension only. One reason is that the model is very local,
so that for the properties we will discuss here, dimensionality is not too important. The
second reason is that comparisons can be made with exact results, which are readily available
only in one dimension. 

Ground state energies can be obtained both by exact diagonalization (on small system
sizes) and by Bethe Ansatz techniques \cite{lieb68,bahder86,marsiglio97}.
We have already outlined in some detail \cite{marsiglio97} the numerical procedure
used to obtain ground state energies for the attractive Hubbard model, and the reader
is referred to those references for further details. 

\section{RESULTS}
\label{sec:res}

\subsection{Ground state energy}
\label{sec:res:ge}

We present results in one dimension and with only the nearest neighbours
included for the electron hopping ($t_\delta\equiv t$).
It has been shown in \refer{marsiglio97} that for large systems,
the grand canonical BCS approximation yields the exact energy in the strong- 
and weak-coupling limits, and in the dilute limit for all coupling strengths.
Thus deviations from the exact results are largest for weak to 
intermediate coupling strengths and
for larger electron density $n\equiv\langle N_e\rangle/N$.  
We therefore anticipate improvements by the canonical BCS approximation 
for these cases, especially for small system size $N$.
Because of the particle-hole symmetry \cite{lieb68,marsiglio97},
we need to study only up to half filling, $n=1$ (i.e., the number of 
pairs is half the number of sites).  
Note that for the exact and canonical calculations, the density is defined 
simply by $n=N_e/N$ for a given number of electrons $N_e$.  
In the results shown below, for changing the density, 
we vary the electron number for a fixed system size, as is the case in the
experiments.

\begin{figure}
\pss{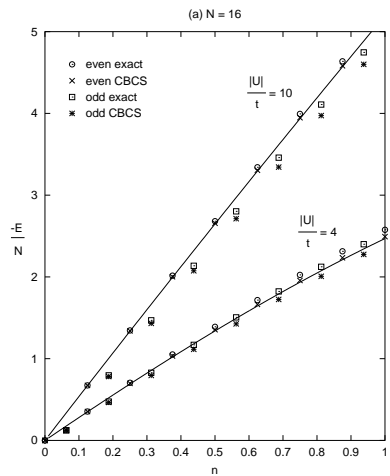}
\caption{(a) Ground state energy as a function of the electron density 
$n=N_e/N$ for 16 sites.  The points are the canonical and exact results
(different symbols for even and odd $N_e$), while the grand canonical ones 
are shown with curves.  The difference between the energies with even and odd
$N_e$ can be seen clearly for $|U|/t=10$.
The canonical BCS results for even $N_e$ are better than those for odd $N_e$, 
and converge to the grand canonical energies for strong coupling.
}
\label{fig1a}
\end{figure}

In Fig.~1, we show the ground state energy per site (absolute value)
as a function of the electron
density for (a) $|U|/t=10$ and 4 and $N=16$, and for (b) $|U|/t=2$ and 
$N=4$, 8 and 32.  The exact and canonical results are plotted with symbols
and the grand canonical ones are shown with curves.  In \fig{fig1a}(a), for
the exact and canonical cases, the energies with even and odd numbers of
electrons are shown in different symbols.  The conventional grand canonical 
BCS wave function contains only the components with even numbers of electrons.
Strictly speaking,
the exact and canonical results for odd numbers of electrons should be
compared with the grand canonical ones in the parity-conserved scheme 
\cite{janko94,braun97} with the odd number parity.  

In \fig{fig1a}(a), we first note the difference between the energies with
even and odd numbers of electrons, as clearly seen for $|U|/t=10$.  The 
difference becomes more apparent for larger coupling strengths, giving rise to 
the even-odd oscillations in the energy as a function of the electron density.
For $|U|/t=10$, we note another difference between the even and odd
electron numbers; the canonical energies are much closer to
the exact ones for the even numbers.  It can also be seen for this strong 
coupling case that the canonical results (for even $N_e$) are converged to the 
grand canonical ones for almost all density values,
and both results are in very good agreement with the exact 
solutions for smaller density.  As the coupling strength becomes smaller,
for a fixed system size,
the canonical energy deviates more from the grand canonical one and,
for an even number of electrons, improves slightly
the agreement with the exact energy for larger density.
This can be seen for $|U|/t=4$ in \fig{fig1a}(a), while the even-odd difference
is now smaller.

\addtocounter{figure}{-1}
\begin{figure}
\psbb{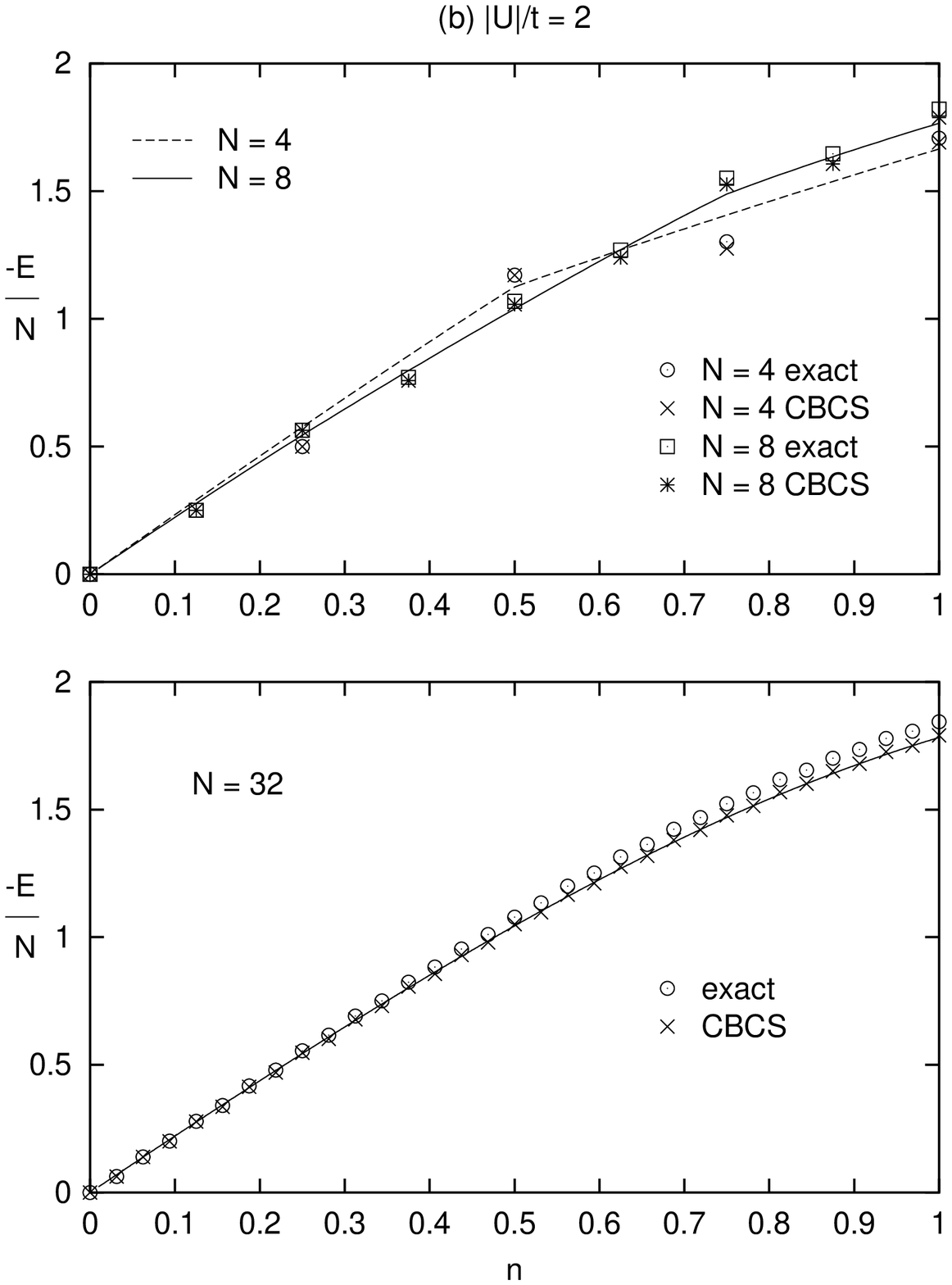}
\caption{(b) Same as (a), but for $|U|/t=2$ and for $N=4$,8 (upper frame)
and 32 (lower frame), and here the even and odd points are not distinguished.
The improvement by the canonical scheme is more apparent for weak coupling
and small system size, as can be seen for $N=4$ and 8.
For $N=32$, the canonical and grand canonical results are more or less 
converged for all densities.
}
\label{fig1b}
\end{figure}

For smaller system size and coupling strength, we see more improvements 
due to the use of the canonical formulation.  In the upper part of \fig{fig1b}(b),
we show the results for $|U|/t=2$ and for $N=4$ and 8.  Note that the 
energy range is magnified compared with \fig{fig1a}(a), and that for the
exact and canonical results, the even and odd energies are not distinguished
by different symbols.  For $N=4$ the
agreement between the canonical and exact energies is excellent for all
values of $n$, and it is still very good for $N=8$. In the latter case,
the grand canonical curve happens to be very close to the exact results
for odd numbers of electrons for larger density.  As explained above, however, 
the grand canonical results shown here must be compared for the even 
numbers only.
In fact, the grand canonical energy for an even number of electrons 
differs most from
the exact and canonical energies for larger density.
On the other hand, the canonical energy converges to the grand canonical 
one as the system size becomes larger, as can be seen in \fig{fig1b}(b)
for $N=32$.  The even-odd difference is negligible for this weak coupling
and large system size.

\begin{figure}
\ps{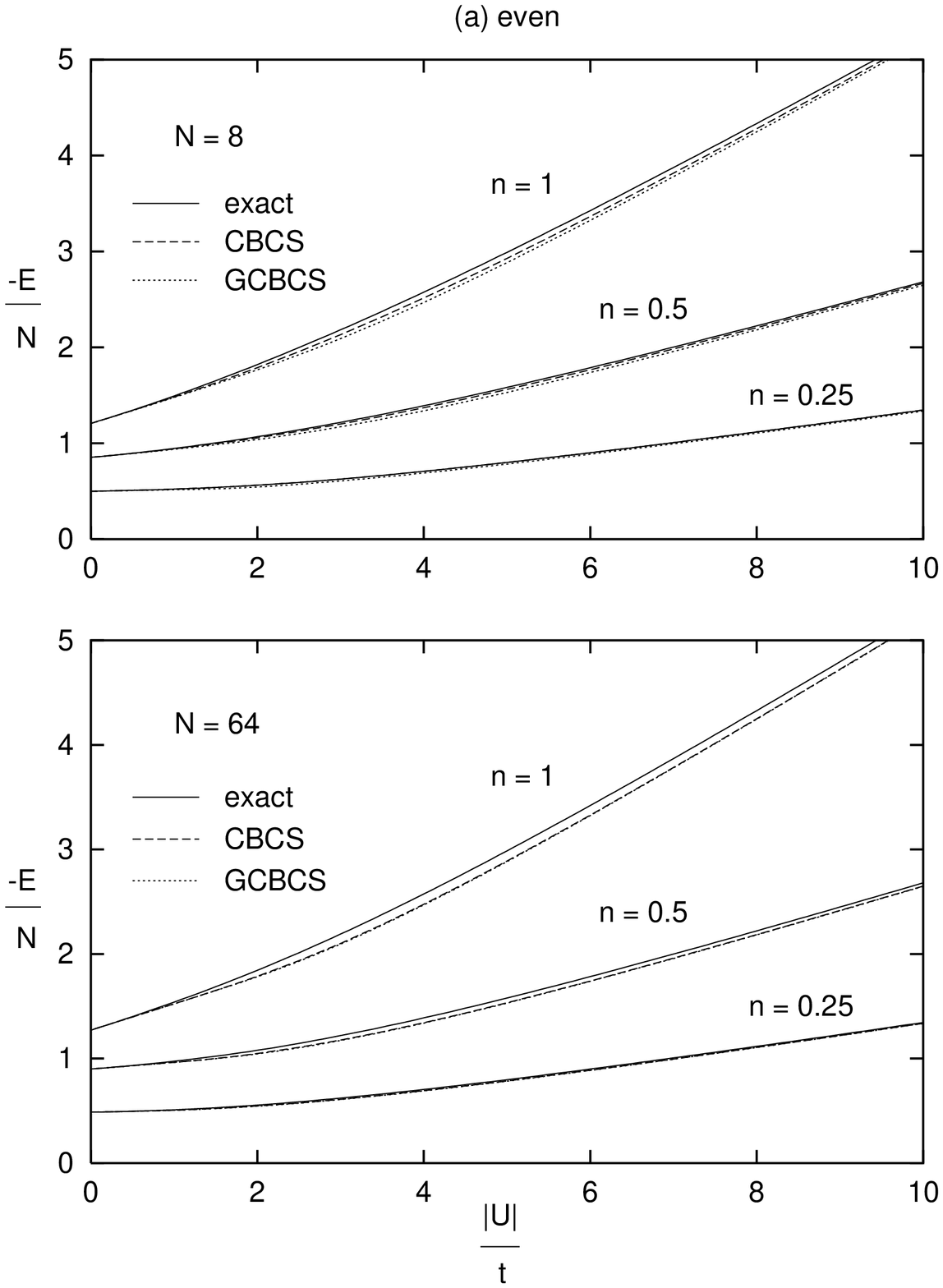}
\caption{(a) Ground state energy as a function of the coupling strength $|U|/t$
for $N=8$ (upper frame) and 64 (lower frame) and for various values of
$n=N_e/N$.  For $N=8$, the improvement by the canonical method can be seen,
and for $n=0.25$, the canonical and exact energies are indistinguishable.  
On the other hand, for $N=64$, the canonical energies are converged to the 
grand canonical ones.  Both the canonical and grand canonical results 
reproduce well the exact solutions for small $n$ and in the zero- and
strong-coupling limit.
}
\label{fig2a}
\end{figure}

In Fig.~2, the ground state energy is plotted as a function of the coupling
strength for $N=8$ and 64, for various densities with (a) even and (b) odd
numbers of electrons.  For small systems, the canonical results improve 
the grand canonical ones, especially for intermediate coupling strengths, 
while the former converge to the latter as the coupling strength is increased.
This can be seen for $N=8$ in \fig{fig2a}(a).  In fact for $n=0.25$, 
the exact (solid) and canonical (dashed) lines are indistinguishable, 
though the grand canonical one deviates from them only slightly.  
The three results converge as the coupling goes to zero, and also in the
strong-coupling limit.
For $N=64$, the size is large enough that the canonical and grand canonical 
results are converged in the given scale 
for all densities and coupling strengths.

\addtocounter{figure}{-1}
\begin{figure}
\ps{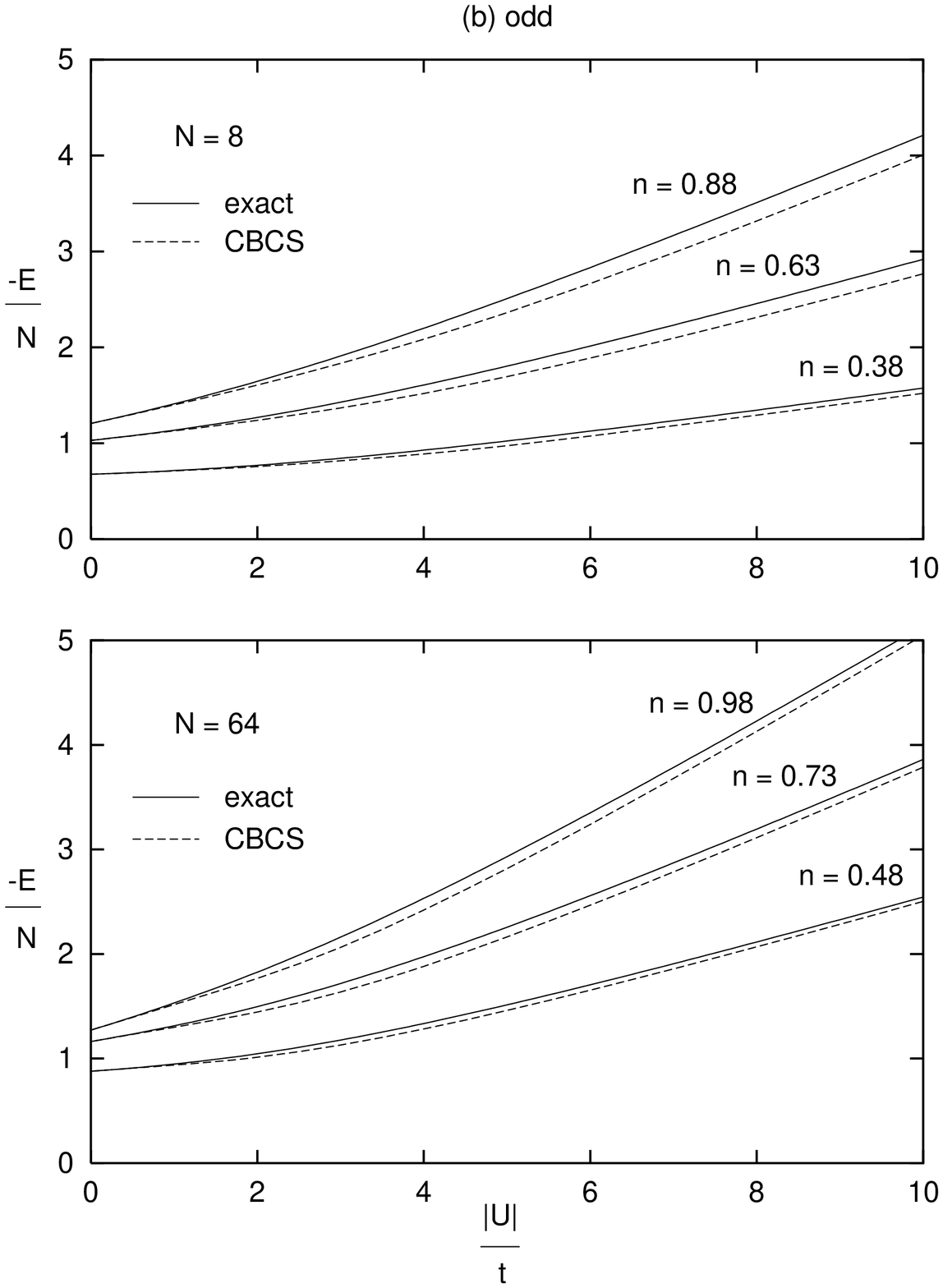}
\caption{(b) Same as (a), but 
here we compare the exact and canonical BCS results for an odd number of electrons $N_e$.
For $N=8$, the energies with odd $N_e$'s are poorly reproduced 
for large $|U|$ by the canonical BCS approximation, compared with the 
even-$N_e$ case shown in (a).  As the system size is increased, the difference
between the errors in the energy for even and odd $N_e$'s becomes smaller,
as can be seen for $N=64$.
}
\label{fig2b}
\end{figure}

In \fig{fig2a}(b), the canonical and exact energies are compared for odd
electron numbers.  For $N=8$, compared with \fig{fig2a}(a), the exact
energy is reproduced rather poorly by the canonical BCS approximation,
especially for larger coupling strengths.
Moreover, unlike the even number case, the canonical BCS results do not 
converge to the exact ones as the coupling strength is increased further.
This difference between the errors in the energy by the even- and odd-$N_e$
canonical BCS approximation turns out to be smaller as the system size becomes
larger.
For $N=64$, the difference between the even and odd cases has diminished 
significantly from the $N=8$ case.

\subsection{Energy gap}
\label{sec:res:gap}

\subsubsection{Finite size effects}
\label{sec:res:finite}

\begin{figure}
\psbb{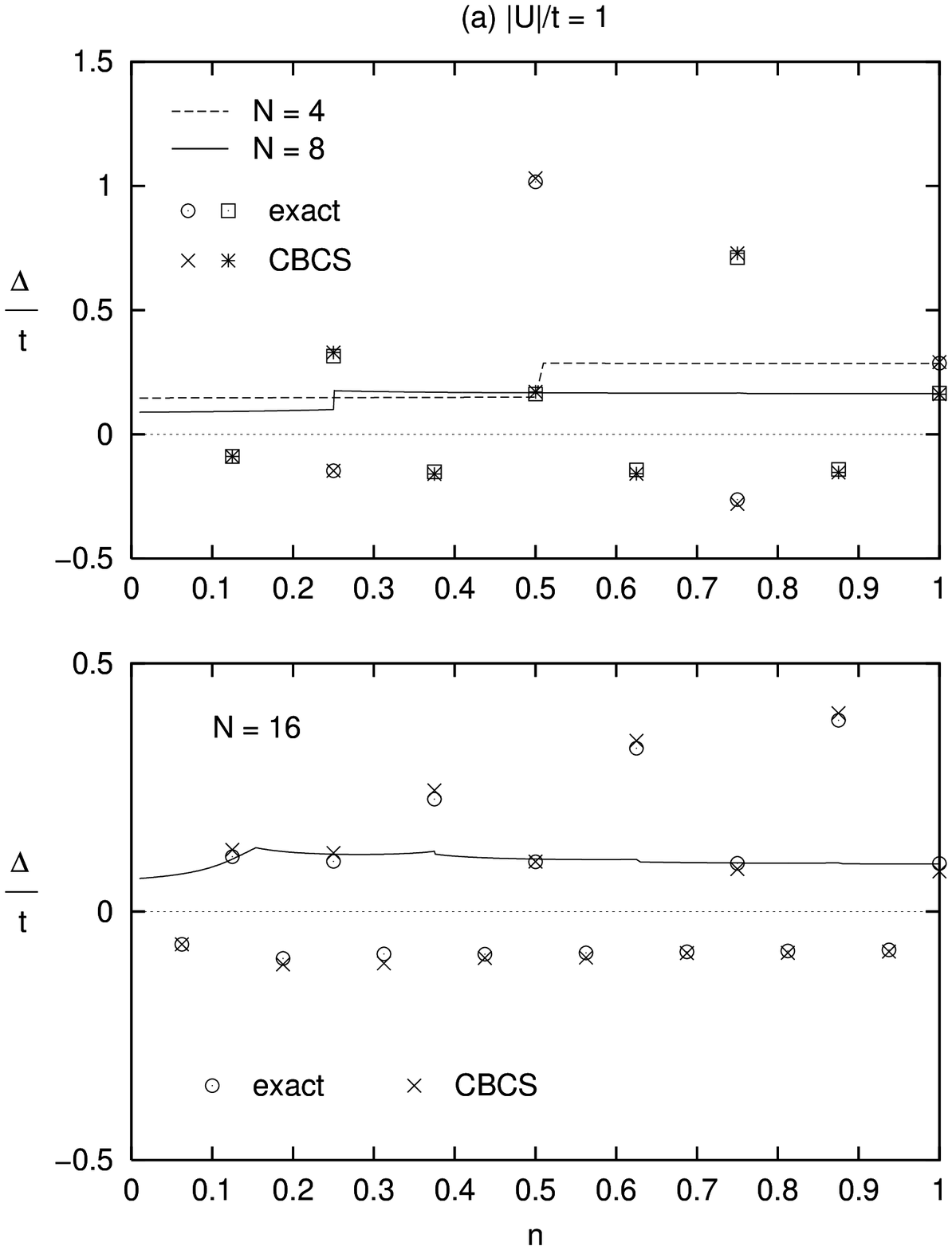}
\caption{(a) Energy gap as a function of the electron density $n=N_e/N$ for
$N=4$,8 (upper frame) and 16 (lower frame) and for $|U|/t=1$.  
In the upper figure, the circles and crosses are the exact and canonical
BCS results for $N=4$, respectively; the squares and stars are for $N=8$.
The grand canonical results are shown with curves.
The gap for odd $N_e$ is negative, and for even $N_e$, the {\it super-even}
oscillations ($N_e=4 m$ vs. $4 m + 2$) can be seen.
The canonical results are in excellent agreement with the exact ones,
while the grand canonical BCS result completely misses the gaps for 
$N_e=4 m + 2$.}
\label{fig3a}
\end{figure}

We show the energy gap as a function of the density
for (a) $|U|/t=1$, (b) $|U|/t=1.5$ and (c) $|U|/t=4$;
for $N=4$ and 8 (upper figures) and $N=16$ (lower figures) in \fig{fig3a} 
and for $N=32$ (upper) and 64 (lower) in \fig{fig4a}.
Again, the exact and canonical results ($\Delta(N_e)$ in \eq{cangap})
are plotted with symbols and the
grand canonical results ($\Delta_{\circ}$ in \eq{mingap}) are shown with curves.
In the upper parts of \fig{fig3a}, the exact solutions for $N=4$ and 
8 are the circles and squares, respectively, and the corresponding
canonical results are the crosses and stars.
The gaps for odd numbers of electrons are negative.  
As discussed above, the conventional BCS wave function that is used in this work
does not contain the odd-$N_e$ components, and thus
the gap parameter defined by \eq{mingap} misses out the gap 
for odd electron numbers.
For even electron numbers, the canonical BCS method improves the
grand canonical results significantly for weak coupling and small system size.

\addtocounter{figure}{-1}
\begin{figure}
\psbb{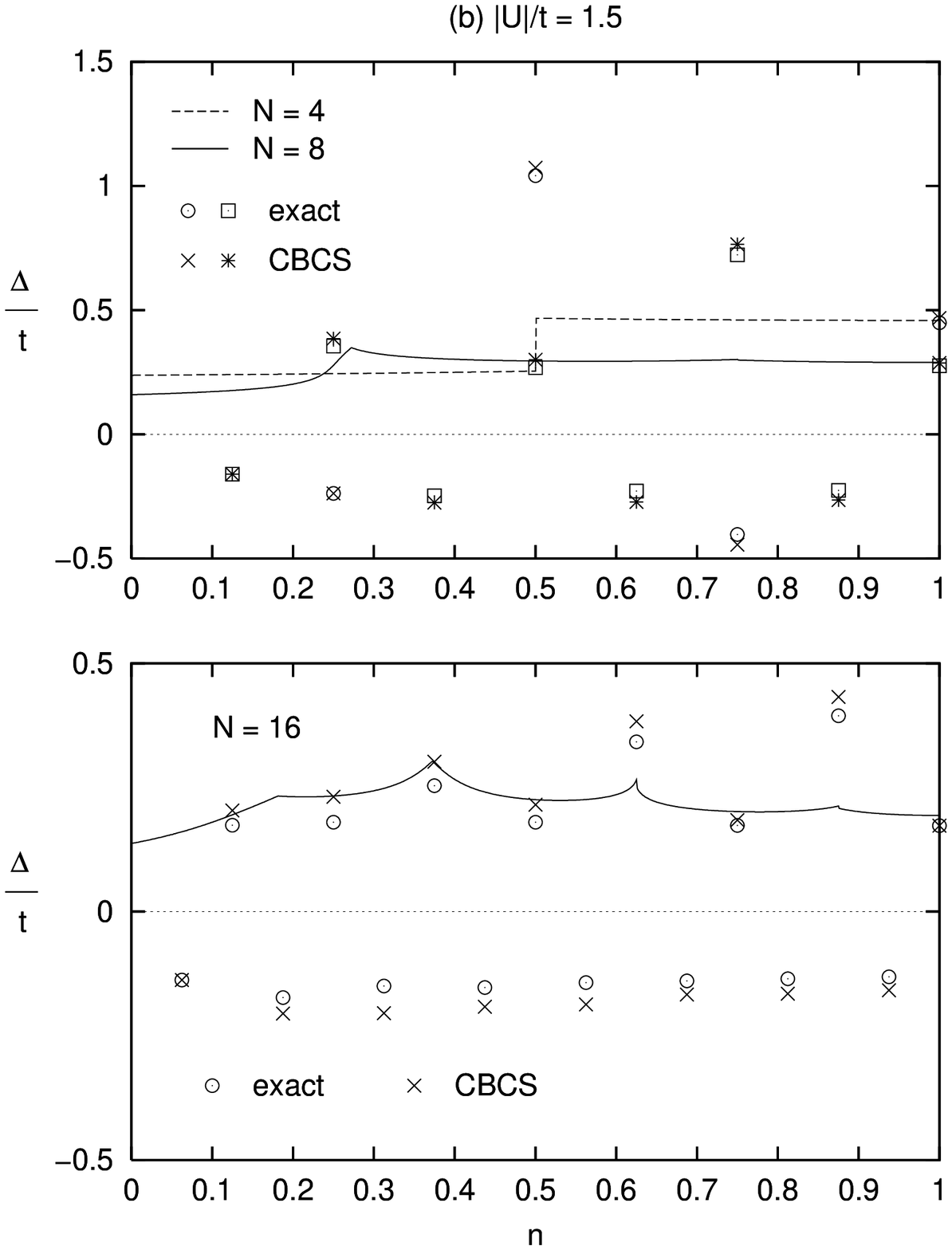}
\caption{(b) Same as (a), but for $|U|/t=1.5$.  The symbols used for the exact
and canonical gaps for $N=4$ and 8 are the same as in (a). 
The canonical results are still in good agreement with the exact ones.}
\label{fig3b}
\end{figure}
\noindent

In \fig{fig3a}(a) for $|U|/t=1$ and $N=4$, 8 and 16, 
we can see the excellent agreement
between the exact and canonical BCS results.  Furthermore, there is a striking feature 
in these figures -- the difference between the gaps for $N_e=4 m$ 
and $N_e=4 m + 2$, where $m$ is an integer.
(Note that for all the results shown,
the number of sites is a multiple of four.)
The gaps for $N_e=4 m + 2$ are much larger than those for $N_e=4 m$.  
Moreover, it is clearly seen for $N=16$ that the gaps for $N_e=4 m + 2$ increase
as the density increases.
We call these oscillations in the gap as a function of the even-$N_e$ 
the {\it super-even} effect: the system is more stable with $4 m + 2$ electrons
than with $4 m$ electrons.  This effect is more pronounced when the coupling is
weak, and it stems from the quantized and doubly degenerate energy levels,
$\epsilon_{\bk}=\epsilon_{-\bk}$, of the unperturbed system.
As will be shown later, when the coupling is weak,
the occupation probabilities of the unperturbed states can be approximated by
those for zero coupling.
We can thus understand the super-even effect as follows, 
in terms of the unperturbed energy levels which are occupied
by pairs of electrons up to the Fermi level and are empty above it.
To simplify the discussion, we ignore the energy change due to the 
blocked states by unpaired electrons.

In the case of one dimension and with $t_\delta=t$, the unperturbed energy,
\eq{kenergy}, reduces to 
$\epsilon_{k}=-2 t\,{\rm cos}\,k R$, where $R$ is the lattice constant.
From the periodic boundary condition, $k R=2\pi j/N$ ($-N/2 < j \leq N/2$), and
each energy level is degenerate for $\pm k R$ except for $k R =0$ and $\pi$.
Hence each level with $0 < |k R| < \pi$ can accommodate two pairs,
($k\uparrow,-k\downarrow$) and ($-k\uparrow,k\downarrow$).
Thus when there are $4 m + 2$ electrons, in the ground state,
all the levels from $k R =0$
up to $k_F R$ are fully occupied by the pairs, while with $4 m$ electrons, 
the Fermi level has a vacancy for one more pair.  
Therefore, when there are $4 m$ electrons, a way to break a pair with the
minimum energy is to flip the spin of an electron of the pair at the
Fermi level.  The unpaired electrons then occupy the states $\pm k_F$
and there is no extra cost for the kinetic energy.
On the other hand, when there are $4 m + 2$ electrons, one cannot break
either pair at the Fermi level simply by flipping a spin, but
an unpaired electron has to move up to the next available level, 
increasing the kinetic energy.
This is why the gap for pair breaking is larger for $4 m + 2$ electrons
than for $4 m$ electrons.  The fact that the former gap increases as 
a function of $N_e$ is particular to the one-dimensional band structure 
that we use:
the level spacing becomes maximum around $k R =\pi/2$ (i.e., half filling)
and so does the kinetic energy cost for breaking a pair.

\addtocounter{figure}{-1}
\begin{figure}
\psbb{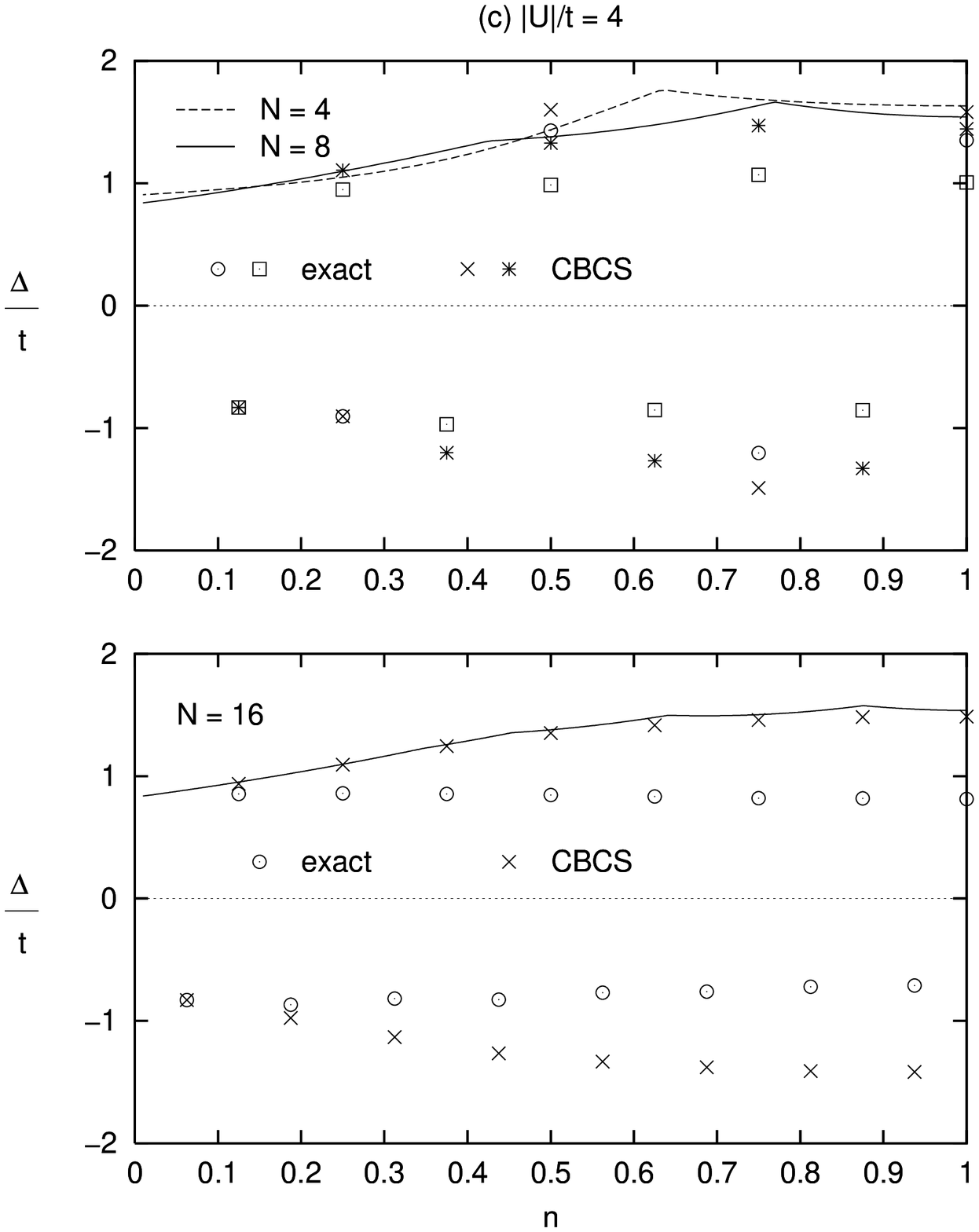}
\caption{(c) Same as (a), but for $|U|/t=4$. For $N=16$ the canonical gaps 
are converged to the grand canonical curve for almost all densities.
Even for $N=4$ and 8, the canonical results are closer to the grand 
canonical ones, and the super-even oscillations have disappeared.}
\label{fig3c}
\end{figure}

The super-even effect can be recognized more clearly 
in \fig{fig4a}(a) for $N=32$ and 64.
The overall scale of the gap becomes smaller for larger system size
(note the reduced scale compared to \fig{fig3a}(a)).
For 32 sites, the canonical BCS results still follow the exact solutions 
closely for almost all density values.
For 64 sites, the system is so large that even for this weak coupling,
the canonical gaps are converged to the grand canonical curve
in the dilute limit.  The canonical results more or less follow
the exact ones for $n\agt 0.3$, where the super-even effect is manifest.

\begin{figure}
\psbb{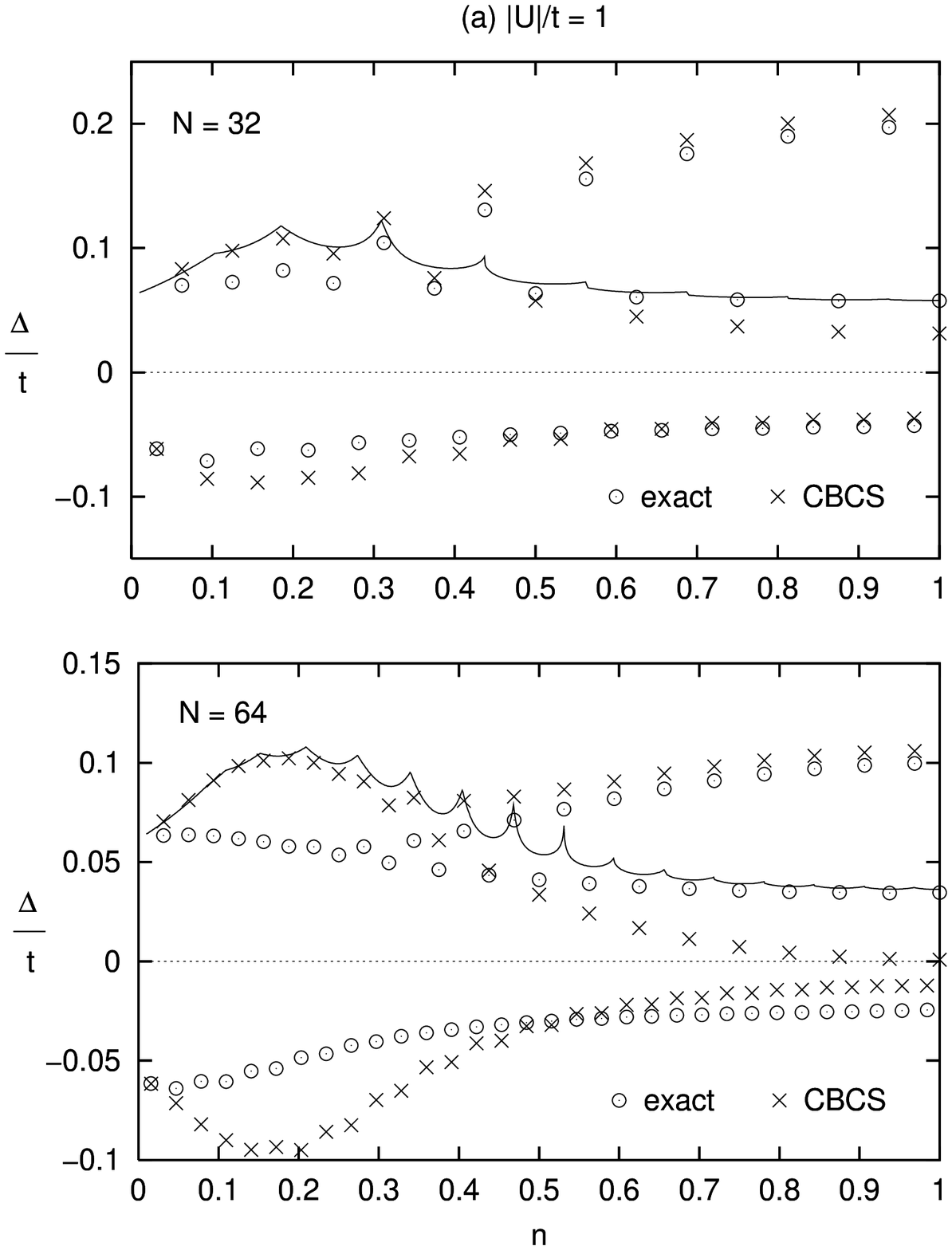}
\caption{(a) Same as Fig.3(a), but for larger system size; 
$N=32$ (upper frame) and $N=64$ (lower frame).  The super-even oscillations
can be seen clearly, whereas the grand canonical BCS misses the $4 m + 2$ gaps
completely.  While for N=32 the canonical BCS results still follow
the exact ones closely, for N=64 the former are closer to the grand 
canonical curve for low electron density.}
\label{fig4a}
\end{figure}

It can be seen in Figs.~3(a) and 4(a) that 
for larger density, where the $4 m$ and $4 m + 2$ gaps are well separated,
the grand canonical BCS completely misses the exact gaps for $N_e=4 m + 2$.
On the other hand, the grand canonical curves appear to 
follow closely the exact $4 m$ gaps.
However, as the size is made larger than 64 sites, 
the exact $4 m$ gaps become smaller than the gap given by the grand 
canonical curve
(see the bottom graph of \fig{fig4a}(a))
and finally in the bulk limit, they are quite small compared to the grand 
canonical values; at $n=1.0$, the exact and grand canonical $\Delta/t$ are
about 0.003 and 0.015, respectively \cite{marsiglio97}.
Meanwhile, the canonical $4 m$ gaps go up towards the grand canonical
curve, as they (and the $4 m +2$ gaps) converge to the grand canonical values.
Hence for some particular sizes (larger than 64 sites), 
the canonical results for the $4 m$ gaps will be closer to the exact ones.

In fact the grand canonical gaps tend to have a discontinuity at $N_e=4 m + 2$, but either value is generally well below the exact or canonical value.
For smaller density where the super-even oscillations are not so prominent,
the grand canonical gaps have cusps at $N_e=4 m + 2$, as can be seen, e.g.,
for $N=64$ and $n\alt 0.5$ in \fig{fig4a}(a).
Also, as the coupling strength is made a little larger but still small
($|U|/t \alt 2$), these discontinuities at larger density are replaced
by cusps.
This can be seen in Figs.~3(b) and 4(b) for $|U|/t=1.5$.
While for the small sizes shown in \fig{fig3b}(b) there are still 
discontinuities for $n\ge 0.5$, for the larger sizes in \fig{fig4b}(b)
the curves are continuous for all densities, with cusps at $N_e=4 m + 2$.
It is intriguing that the grand canonical BCS partially reproduces 
the super-even oscillations in this way.
The grand canonical solutions in the zero-coupling limit will be discussed
further below.

\addtocounter{figure}{-1}
\begin{figure}
\psbb{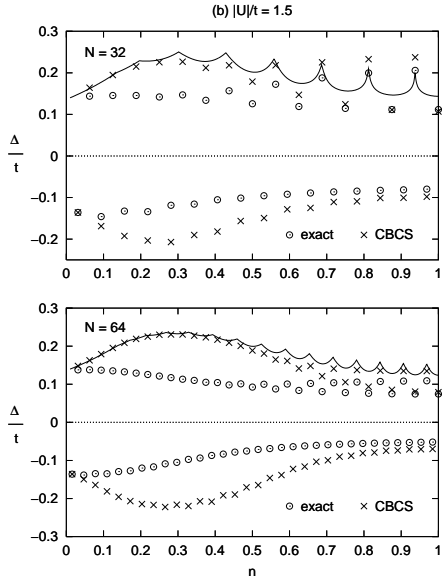}
\caption{(b) Same as (a), but for $|U|/t=1.5$.  The canonical gaps are 
converged to the grand canonical ones for smaller density, while the former
improve the latter still significantly near half filling.}
\label{fig4b}
\end{figure}
\noindent

For $N=4$ and 8 in Fig.~3(b), the agreement between the canonical
and exact gaps is still very good, and the overall structure of the gaps
as a function of the density is the same as for $|U|/t=1$.
For 16 sites, the canonical results deviate slightly but still reproduce 
the exact gaps well.  For small density, the canonical even-$N_e$ gaps are 
closer to the grand canonical curve, while for larger density 
the canonical $N_e=4 m + 2$ gaps significantly 
improve the grand canonical curve.
As the system becomes larger, the super-even oscillations diminish in
amplitudes and also are confined more towards half filling.
Also, as the size increases, for larger density where the 
(exact and canonical) $4 m$ and $4 m + 2$ gaps are well separated, 
the grand canonical curve shifts up relative to them, from near the $4 m$ gaps
to above the $4 m + 2$ gaps.  This can be seen in Figs.~3(b) and 4(b)
in increasing order in size.
For 32 sites, the last two cusps 
in the grand canonical curve happen to be closer to the exact values than
the canonical ones.  However, the canonical gaps capture 
the correct behaviour of the super-even oscillations.
Finally for 64 sites, the canonical gaps are converged to the grand canonical
one for low density, while they are much better for higher density,
especially for $N_e=4 m$.

As the coupling strength is increased further (for a fixed size),
the scale of the gap increases as a whole
for both even and odd $N_e$ (thus the even-odd difference becomes larger),
while the $4 m$ vs. $4 m + 2$ difference decreases.
For example, it can be seen in Figs.~3(c) and 4(c) that $|U|/t=4$ is strong enough 
for most of the sizes shown for the system to reach the ``bulk'' limit,
where the canonical and grand canonical methods hardly differ from one another.
Even for $N=4$ and 8, the super-even structure is gone, and 
the canonical gaps are closer to the grand canonical ones.

\addtocounter{figure}{-1}
\begin{figure}
\psbb{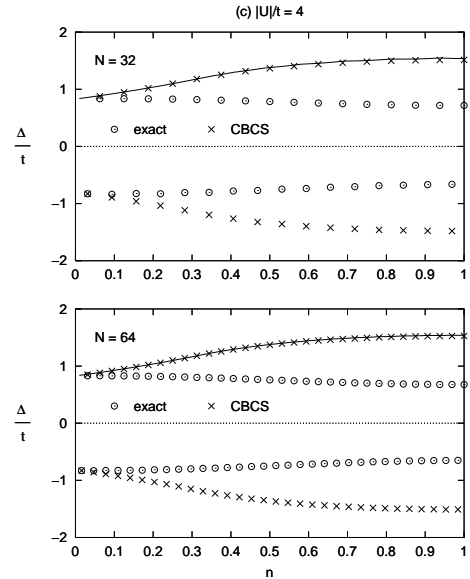}
\caption{(c) Same as (a), but for $|U|/t=4$.  For the large sizes shown,
this is strong enough coupling so that the canonical and grand canonical 
results are converged for all densities, and the super-even oscillations
have disappeared.}
\label{fig4c}
\end{figure}

\begin{figure}
\ps{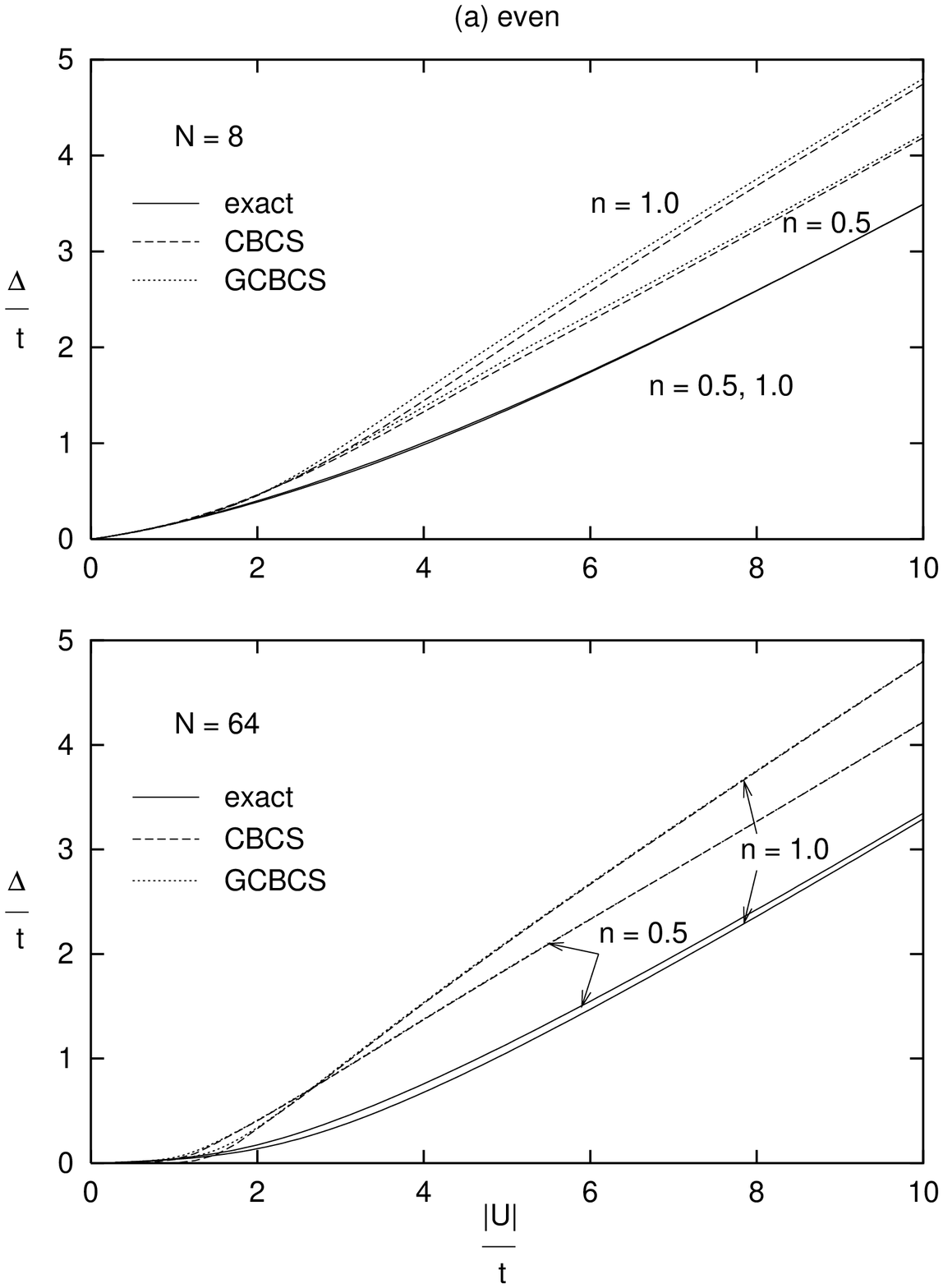}
\caption{(a) Energy gap as a function of the coupling strength $|U|/t$
for $N=8$ (upper frame) and 64 (lower frame), for the densities
$n=0.5$ and 1.0.  For $N=8$ (and for $N_e=4 m$)
the canonical and grand canonical results are equally good for weak coupling,
while the former improve the latter slightly for stronger coupling.
For $N=64$ the canonical and grand canonical results can hardly be 
distinguished for almost all the coupling strengths shown.}
\label{fig5a}
\end{figure}

To see the effect of the coupling strength on the energy gap, we plot in
\fig{fig5a}(a) the gap as a function of the coupling strength
for $N=8$ (upper) and 64 (lower), for quarter ($n=0.5$) and half filling.
Interestingly, for $N=8$ the exact gaps for quarter and half filling
are almost the same for all $|U|$ values, whereas for $N=64$ they are
somewhat different.  
On the contrary, in the BCS picture (both canonical and grand canonical),
the difference between quarter and half filling in the strong-coupling
limit is about the same for small ($N=8$) and large ($N=64$) sizes
and much larger than the exact result.
As for the difference between the canonical and grand canonical results,
we first note that $N_e=4 m$ for both cases shown in \fig{fig5a}(a).
For $N=8$, the canonical and grand canonical results are equally good for
weak coupling ($|U|/t\alt 2$), whereas the canonical gaps improve the grand
canonical ones slightly for stronger coupling.
For large systems and for very weak coupling, 
the grand canonical results are better than the canonical ones.
This can be seen for $N=64$ for $|U|/t \alt 1$ for half filling,
as we have seen in \fig{fig4a}(a).
As the coupling becomes stronger, the canonical gap converges to the grand
canonical one, and this happens faster for larger systems.
Indeed for 64 sites, 
the two curves for $|U|/t \agt 2$ can barely be distinguished 
in the given scale both for quarter and half filling,
although half filling is a special case
where the canonical gap defined by \eq{cangap} does not converge
to the conventional BCS gap -- this will be discussed shortly.

In \fig{fig5b}(b), we compare the magnitude of the exact and canonical 
gaps for even and odd $N_e$ for half filling, for 8 and 64 sites.
For smaller systems, the magnitude of the gap for even $N_e$ ($=4 m$)
is a little larger than the one for odd $N_e$ ($=4 m \pm 1$), as can be seen
for $N=8$.
This difference is slightly larger for the exact solutions 
than the canonical BCS results.
As the coupling strength or the system size is increased, this difference
between the magnitude of the even ($4 m$) and odd gaps diminishes.
This can be seen in \fig{fig5b}(b).
Also note that the difference between the BCS gap and the exact gap 
increases as the system size
increases. (This can be seen in \fig{fig5a}(a) as well.)
This is due to the fact that the exact gap becomes smaller for larger systems,
whereas the canonical gap in the strong-coupling limit hardly changes 
as the system becomes larger (and $N=64$ is large enough to be the bulk limit
for $|U|/t\agt 4$).

\addtocounter{figure}{-1}
\begin{figure}
\ps{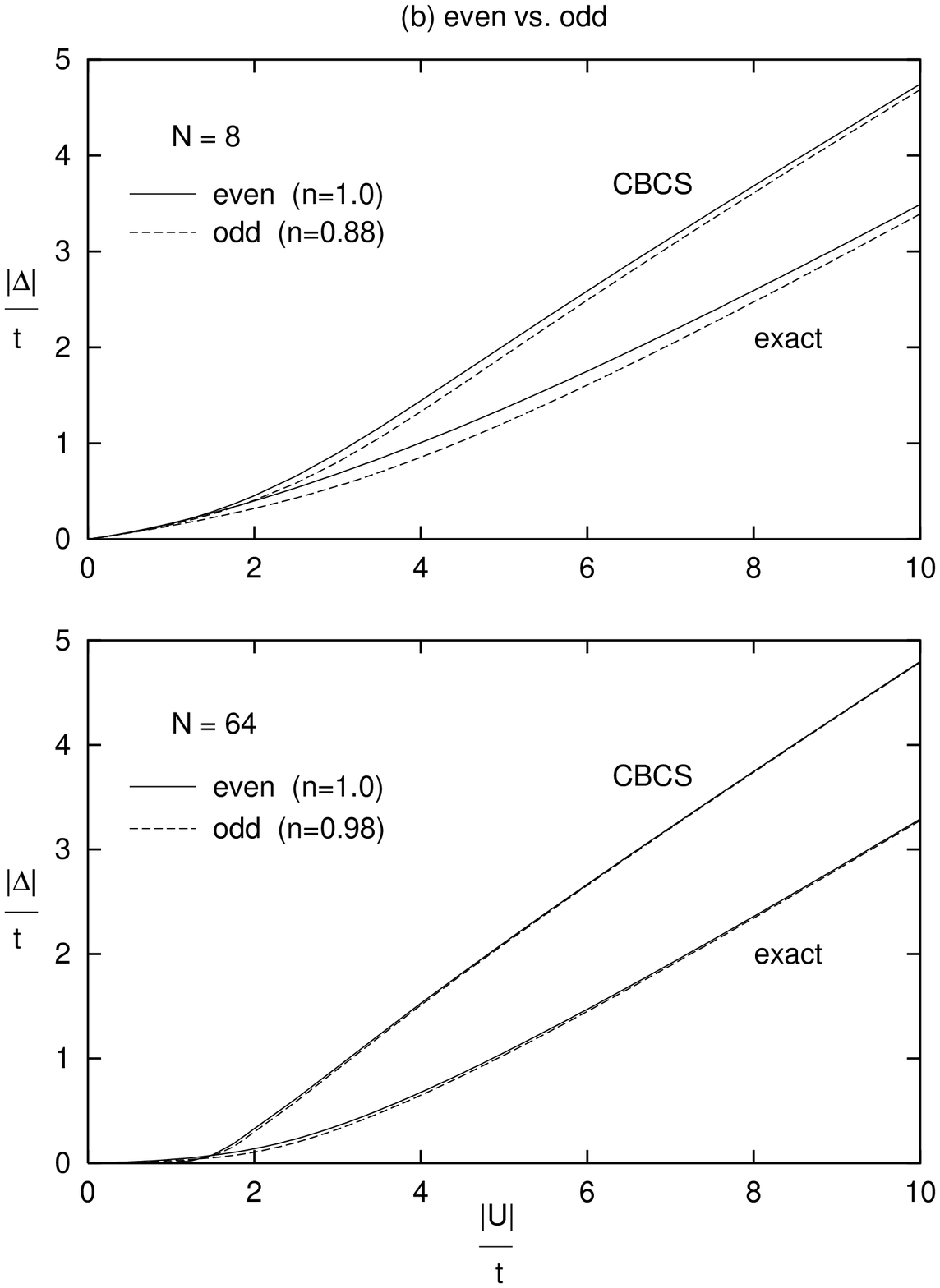}
\caption{(b) Same as (a), but here the magnitude of the gaps for even and
odd electron numbers are compared for the exact and canonical BCS results.
The comparison is made at half filling, so that even $N_e=N$ and odd 
$N_e=N-1$.  For $N=8$ the even gaps are slightly larger in magnitude than 
the odd ones, while for $N=64$ the difference has diminished.}
\label{fig5b}
\end{figure}

\subsubsection{Grand canonical gap for weak coupling}
\label{sec:res:zero}

\begin{figure}
\ps{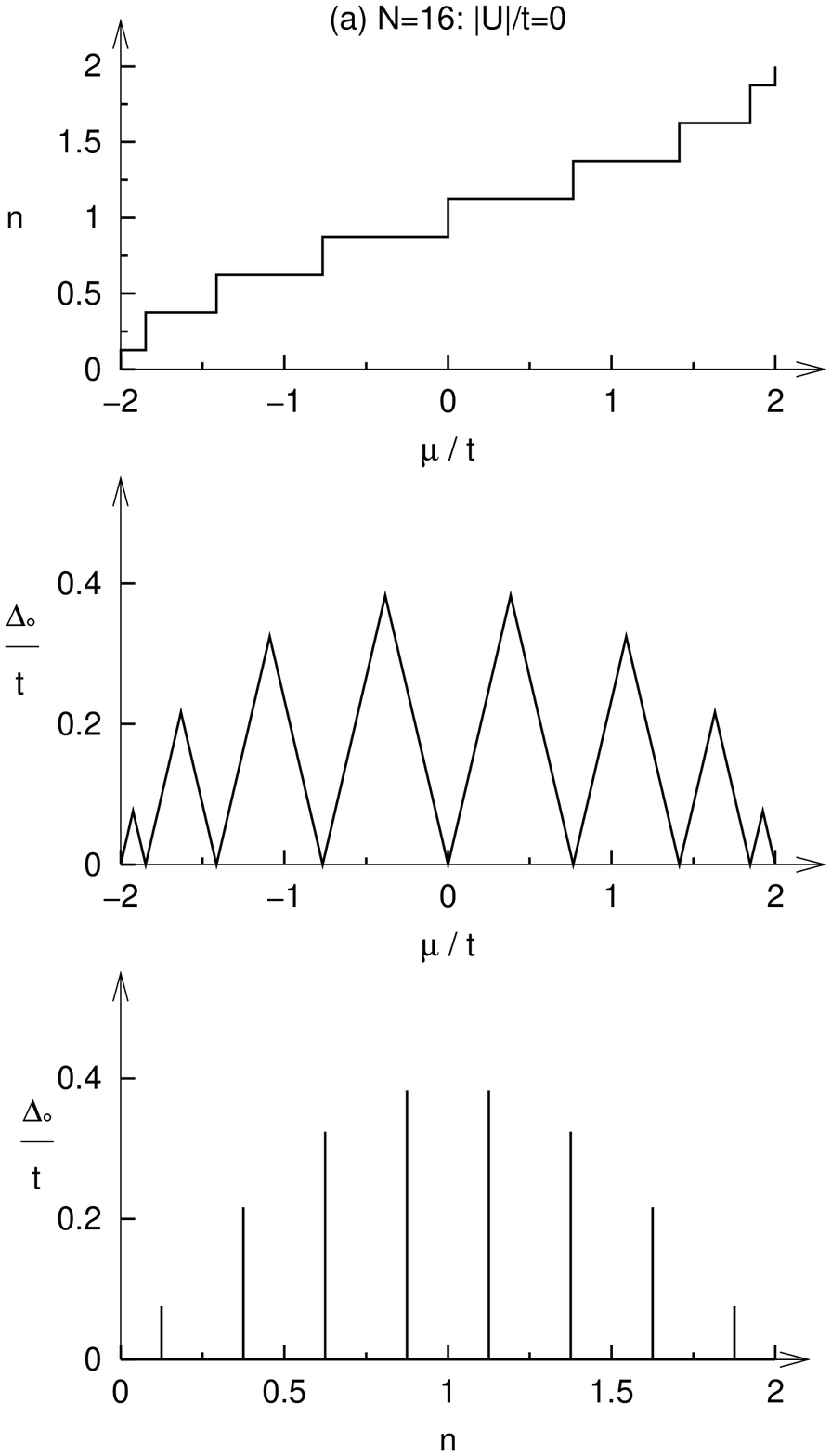}
\caption{(a) The electron density $n$ (top frame) and the gap $\Delta_\circ$
(middle frame) as a function of the chemical potential $\mu$, and 
$\Delta_\circ$ as a function of $n$ (bottom frame), for $N=16$ and 
for zero coupling.  The fact that the density is a step function of $\mu$ 
and that the gap exists is because of quantized energy levels due to the finite system size.
The $\Delta_\circ$ has peaks when $|\epsilon_k-\tilde\mu|$ becomes a maximum.}
\label{fig6a}
\end{figure}

As mentioned above, for very weak coupling, the grand canonical gaps have
discontinuities for $N_e=4 m + 2$ at larger density.
We can understand how these discontinuities arise, by looking at the density
as a function of the chemical potential in the zero-coupling limit.
In \fig{fig6a}(a) we show the density (top) and the gap $\Delta_\circ$ 
(middle) as a function of the chemical potential $\mu$, and 
$\Delta_\circ$ as a function of the density $n$ (bottom), for zero coupling
and $N=16$.
When $|U|=0$, $\Delta_{\rm BCS}=0$, $\tilde\mu=\mu$ and 
$E_k=|\epsilon_k-\mu|$.
There is no gap equation and $n$ is simply determined by \eq{bcs_num} 
for a given $\mu$.
Thus the existence of a gap is solely due to the finite system size.
It can be seen in the top figure that
as $\mu$ changes continuously, the density $n$ changes as a step function:
it is multi-valued when $\mu$ is equal to any of the discrete $\epsilon_k$ 
due to the orbital and spin degeneracy.
As $\mu$ moves from one level to the next, the density stays the same, whereas
$\Delta_\circ={\rm min} (E_{k})$ increases from zero, peaks when 
$\mu$ is precisely between the two levels, and falls to zero again.
Hence $\Delta_\circ$ as a function of $n$ has $\delta$-function-like peaks,
as seen in the bottom figure.

When the coupling is nonzero, $n$ and $\mu$ must satisfy
not only the number equation \eqn{bcs_num} but the gap equation
\be
{1\over |U|}={1\over N}\,\sum_{\bf k}\,{1\over 2\,E_{\bf k}}
={1\over N}\,\sum_{\bf k}\,{1\over 2\,\sqrt{(\epsilon_{\bf k} - \tilde\mu)^2 + \Delta_{\rm BCS}^2}}\;,\label{bcs_gap}
\ee
which is obtained by substituting \eq{soln} into \eq{pairpot}.
It turns out that when the coupling is weak, $\tilde \mu$ given by
\eq{hart} cannot take certain values in between two levels
(corresponding to the plateau regions in $n$ in \fig{fig6a}(a)), and this 
can be understood in the following way.

For weak coupling, we can consider the relation between the density and
chemical potential ($\tilde\mu$) roughly in the same way as for the
zero-coupling case:
for $N_e=4 m$, $\tilde\mu$ is very close to one of the levels $\epsilon_k$ 
(when $|U|=0$, it is equal to $\epsilon_k$), while for $N_e=4 m + 2$ it must
be in between two levels.  Therefore for $N_e=4 m$,
$\epsilon_k-\tilde\mu\simeq 0$ and thus $\Delta_{\rm BCS}$ 
must be finite for $n$ to have a certain value (see \eq{bcs_num}).
For $N_e=4 m + 2$, however, if $\epsilon_k-\tilde\mu$ is finite, then
$\Delta_{\rm BCS}$ is driven to zero so that the RHS of \eq{bcs_gap} attains
a large enough value for small $|U|$.
Indeed, for sufficiently small $|U|$, even if $\Delta_{\rm BCS}$ is zero,
\eq{bcs_gap} {\it cannot} be satisfied for $\tilde\mu$ near the middle of
the range between two levels.  Instead $\tilde\mu$ is driven towards either 
level, so that the RHS of \eq{bcs_gap} increases sufficiently to equal the LHS.
The number equation \eqn{bcs_num} is still satisfied, since 
$\Delta_{\rm BCS}\simeq 0$.  Thus certain values of $\tilde\mu$ are not allowed
for small values of $|U|$.
This situation is illustrated in \fig{fig6b}(b) for $|U|/t=1$.
In the top part, the abrupt jumps in $n$ seen in \fig{fig6a}(a) have been
somewhat smoothed, whereas the plateau parts have disappeared.
Accordingly the gap becomes discontinuous as a function of $\tilde\mu$ 
as well as $n$.
In the latter case, $\Delta_\circ$ indeed has two values for a given $n$
at each discontinuity shown, corresponding to the two solutions for $\tilde\mu$
for coming up from a lower level and coming down from the next level.
In any case, all values of $n$ are possible (in contrast to $\tilde\mu$).

\addtocounter{figure}{-1}
\begin{figure}
\ps{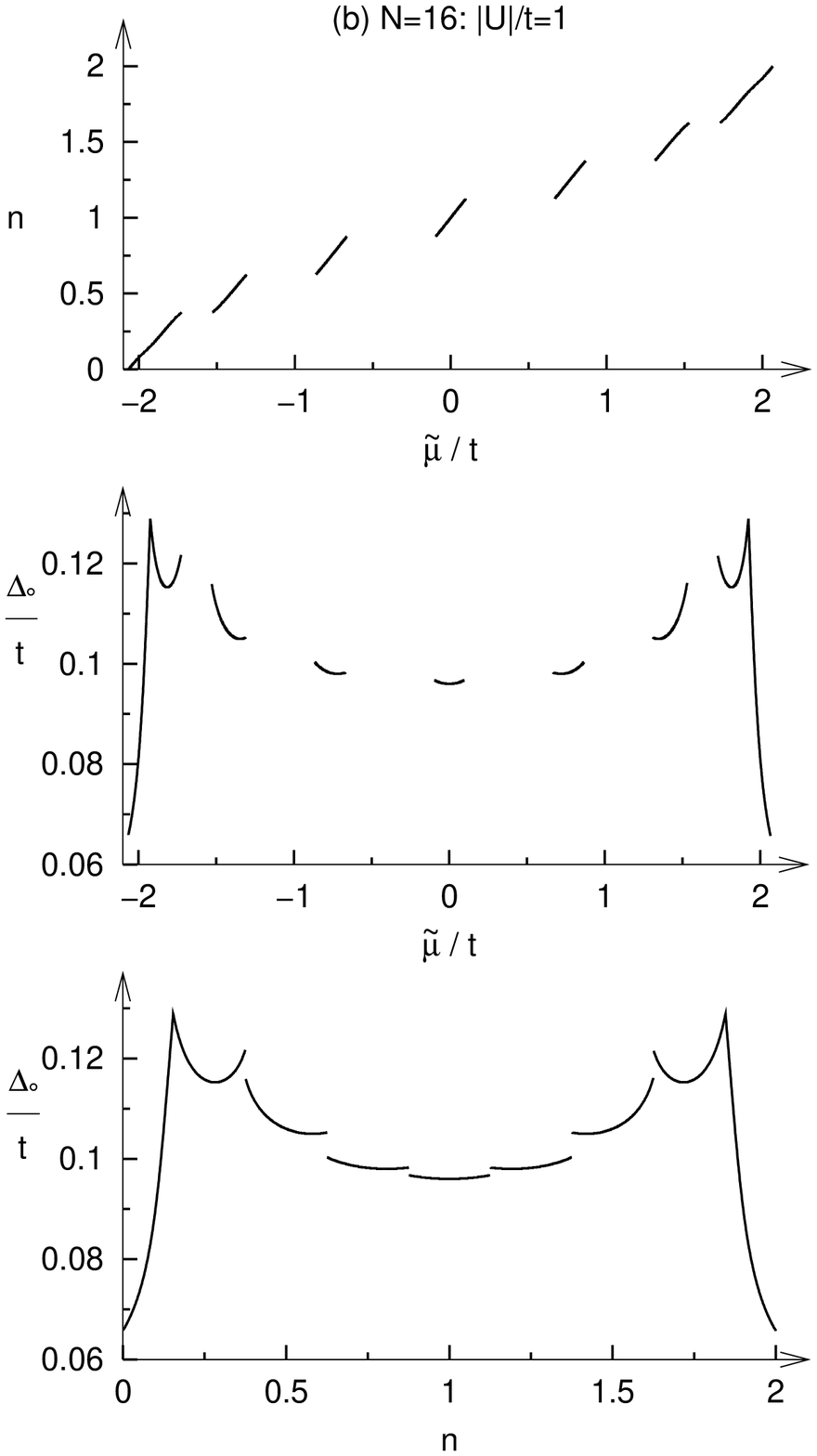}
\caption{(b) Same as (a), but for $|U|/t=1$ and now the ``chemical potential''
is $\tilde\mu$ given by \eq{hart}.  The $\tilde\mu$ cannot take 
certain values in between unperturbed energy levels, corresponding to the
plateau regions in (a).}
\label{fig6b}
\end{figure}

For smaller density or larger system size, the level spacings become smaller,
and it becomes possible for $\tilde\mu$ to have all the values in between
levels.  This also occurs for stronger coupling, as
can be seen in \fig{fig6c}(c) for $|U|/t=2$, where $\Delta_\circ$ is now 
continuous and has cusps at $N_e=4 m + 2$.
These cusps come about when the coupling is still weak enough so that 
$\Delta_{\rm BCS}$ is small, and 
for the same reason as in the zero-coupling case, that is,
when $|\epsilon_k-\tilde\mu|$ becomes a maximum.

\addtocounter{figure}{-1}
\ps{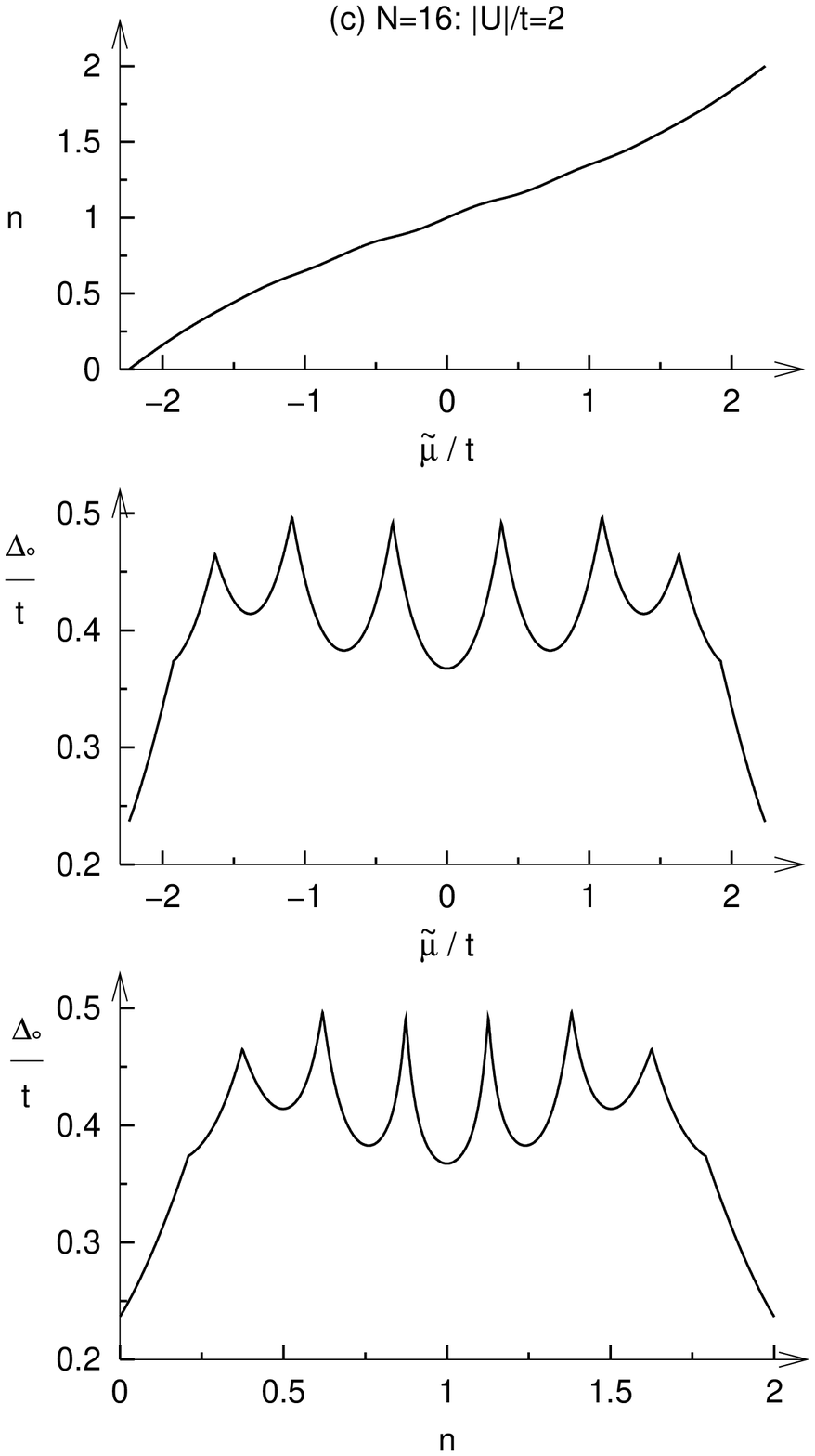}
\begin{figure}
\caption{(c) Same as (a), but for $|U|/t=2$.  This coupling strength is large 
enough so that $\tilde\mu$ can take all the values in between levels. On the 
other hand, the coupling is weak enough so that $\Delta_{\rm BCS}$ is small, 
and $\Delta_\circ$ has cusps when $|\epsilon_k-\tilde\mu|$ becomes a maximum, 
similarly to the zero-coupling case.}
\label{fig6c}
\end{figure}

It is intriguing that the weak but nonzero coupling picture
described above does not interpolate smoothly to the zero-coupling case.
As $|U|$ decreases, the slope of $n$ (centred around $N_e=4 m$)
as a function of $\tilde\mu$ seen in \fig{fig6b}(b) (top frame) 
will become sharper, while the corresponding curves in $\Delta_\circ(\tilde\mu)$
will be reduced to points.
At exactly $|U|=0$, $n$ becomes vertical around $N_e=4 m$ for all $m$.
However, all values of $\tilde\mu$ in between are allowed now,
and $\Delta_\circ(\tilde\mu)$ becomes continuous as in \fig{fig6a}(a)
(middle frame).

\subsubsection{Quasi-particle energy}
\label{sec:res:bcsgap}

In the conventional BCS theory, there is no distinction between a particle-
and a hole-excitation in that the energy 
$E_{\bf k}=\sqrt{(\epsilon_{\bf k} - \tilde\mu)^2 + \Delta_{\rm BCS}^2}$ is the same 
for either excitation \cite{schrieffer64}.
As defined by \eq{mingap},
the gap $\Delta_{\circ}$ is the lowest quasi-particle energy $E_{\bf k}$.
The minimum energy required for breaking a pair is $2 \Delta_{\circ}$
and thus, the momenta $|\bf k|$ carried by an electron and a hole are the same.
In the canonical BCS scheme, for even $N_e$, the minimum pair-breaking 
energy is given by $2 \Delta(N_e)$ in \eq{cangap}, where the lowest energy is 
chosen for each of the systems with $N_e-1$, $N_e$, and $N_e+1$ electrons.
For the lowest $E_{N_e+1}$ and $E_{N_e-1}$, the momentum carried by 
an unpaired electron and that by a hole, respectively, are not necessarily 
the same.
(This is also the case for the exact solutions
by the Bethe ansatz for a finite system.) 
In this section, we evaluate the gap $\Delta(N_e)$ by taking the same momentum 
for an electron and a hole as in the grand canonical picture, and
compare it with the grand canonical quasi-particle energy.
We should remark that the difference in gaps defined in these two different
ways is generally quite small, but noticeable in certain extreme limits.

\begin{figure}
\ps{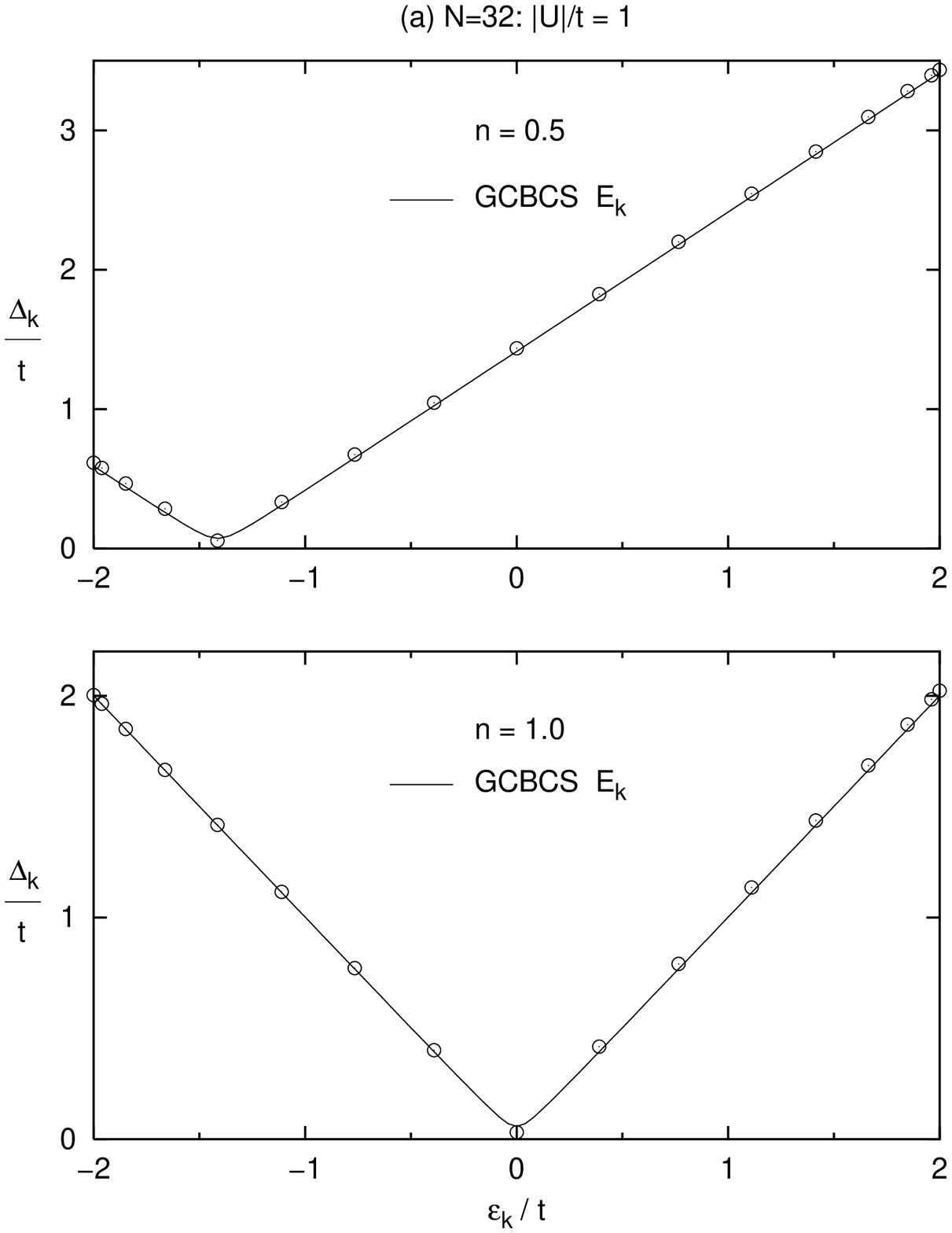}
\caption{(a) The momentum-dependent canonical gap $\Delta_k$, for which
the same $|k|$ is taken for an electron and a hole (circles), and the
grand canonical quasi-particle energy $E_k$ (curves), as a function of the
kinetic energy $\epsilon_k$.  The results are for $N=32$ and $|U|/t=1$,
and for quarter (upper frame) and half (lower frame) filling. For this weak
coupling, $E_k\simeq |\epsilon_k-\tilde\mu|$.}
\label{fig7a}
\end{figure}

In \fig{fig7a}, we plot $\Delta(N_e)$ obtained by the canonical BCS method,
for which the same $|k|$ has been taken for an electron and a hole
and which we call $\Delta_k$,
as a function of the kinetic energy $\epsilon_k$.
The energy band in one dimension is from $-2 t$ to $2 t$, corresponding to
$0 \le |k R| \le \pi$.
We show results for $N=32$ and for quarter and half filling, i.e., 
$N_e=16$ and $N_e=32$, respectively;
and for (a) $|U|/t=1$, (b) $|U|/t=5$ and (c) $|U|/t=50$.
The canonical results are shown with circles, while the grand canonical energy
$E_k=\sqrt{(\epsilon_k - \tilde\mu)^2 + \Delta_{\rm BCS}^2}$
is plotted with solid curves.

We have seen in \fig{fig4a}(a) that for $|U|/t=1$ and $n\agt 0.5$,
the grand canonical gap ($\Delta_{\circ}$) happens to be very close to the 
exact one for $N_e=4 m$, and for $N=32$, also to the canonical gap
(note that $N_e=4 m$ for both quarter and half filling).
We also showed in \fig{fig4a}(c) that $N=32$ is large enough
for the canonical gap to converge to the grand canonical BCS gap
for almost the entire density range, already for $|U|/t=4$.
We find the result that the momentum-dependent canonical gap $\Delta_k$
defined in this section agrees remarkably well with the quasi-particle
energy $E_k$ (defined in the grand canonical context) for 
{\it all quasi-particle momenta} (i.e., not just at the minimum in 
Figs.~7(a)-(c)).

In the canonical picture,
when the coupling is weak and there are $4 m$ electrons,
the lowest energy for breaking a pair is to create an electron and a hole
both at the Fermi level.
Thus $|k|$ for an electron and a hole is the same, and this explains 
the super-even effect.  
The gap with this configuration is 
the minimum value seen in \fig{fig7a}(a) for both quarter and half filling.
In the grand canonical scheme, $\Delta_{\rm BCS}$ becomes very small
for this weak coupling
and hence $E_k \simeq |\epsilon_k - \tilde\mu|$, where
$\tilde \mu$ turns out to be almost at the Fermi level (slightly above)
for quarter filling and it is at the Fermi level for half filling 
($\tilde \mu\equiv 0$).  

\addtocounter{figure}{-1}
\begin{figure}
\ps{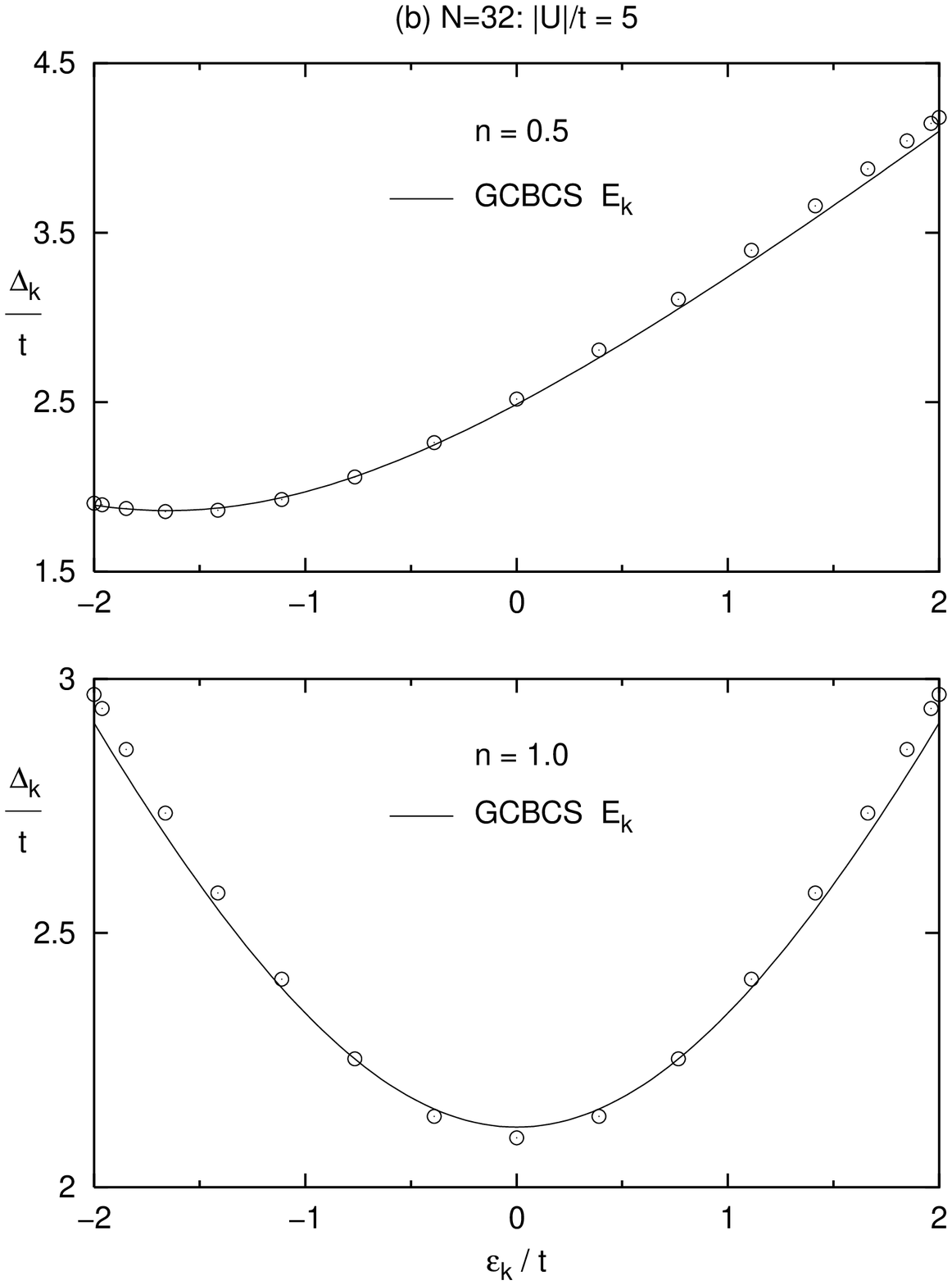}
\caption{(b) Same as (a), but for $|U|/t=5$.}
\label{fig7b}
\end{figure}

As the coupling strength increases,
the momentum of an unpaired electron (or a hole) that yields the lowest energy
for an odd $N_e$ system shifts from the Fermi momentum in the zero-coupling 
limit towards zero momentum -- it eventually reaches zero momentum, that is,
the bottom of the non-interacting band, in the strong-coupling limit.
This happens sooner (i.e., for smaller $|U|$) for smaller number of electrons:
not only the Fermi momentum for zero coupling is closer to zero than
those for larger $N_e$, but also the optimal momentum starts shifting at
weaker coupling.
In the case of quarter filling for $N=32$,
the optimal momenta for $N_e=15$ and 17 are both $|k R|=\pi/4$ for zero coupling,
while for $|U|/t=5$, the ground state energies $E_{15}$ and $E_{17}$ have
different momentum dependence and have a minimum at 
$|k R|=3\pi/16$ and $\pi/4$, respectively.
Yet if we take the same momentum for both $N_e=15$ and 17, 
$\Delta_k$ agrees very well with $E_k$ for all the $|k|$'s, as seen
in \fig{fig7b}(b).
The $\Delta_k$ has its minimum when $E_{15}+E_{17}$ is the smallest; this
occurs at $|k R|=3\pi/16$ for $|U|/t=5$.
For half filling, the optimal momenta for both $N_e=31$ and 33 are still
$|k R|=\pi/2$ ($\epsilon_k=0$) for $|U|/t=5$, as in the zero-coupling case.

The preceding discussion illustrates that to obtain the grand canonical limit,
one must in general choose the quasi-particle momentum (identical for the
particle and hole cases) to minimize the sum of the two odd-electron energies. For
intermediate to strong coupling this will not be given by the momentum expected from
the non-interacting limit (in Ref. \cite{braun98} this procedure worked because
they adopted a particle-hole symmetric model -- see below). The last case (next paragraph)
shows this for a very strong coupling example.

\addtocounter{figure}{-1}
\begin{figure}
\ps{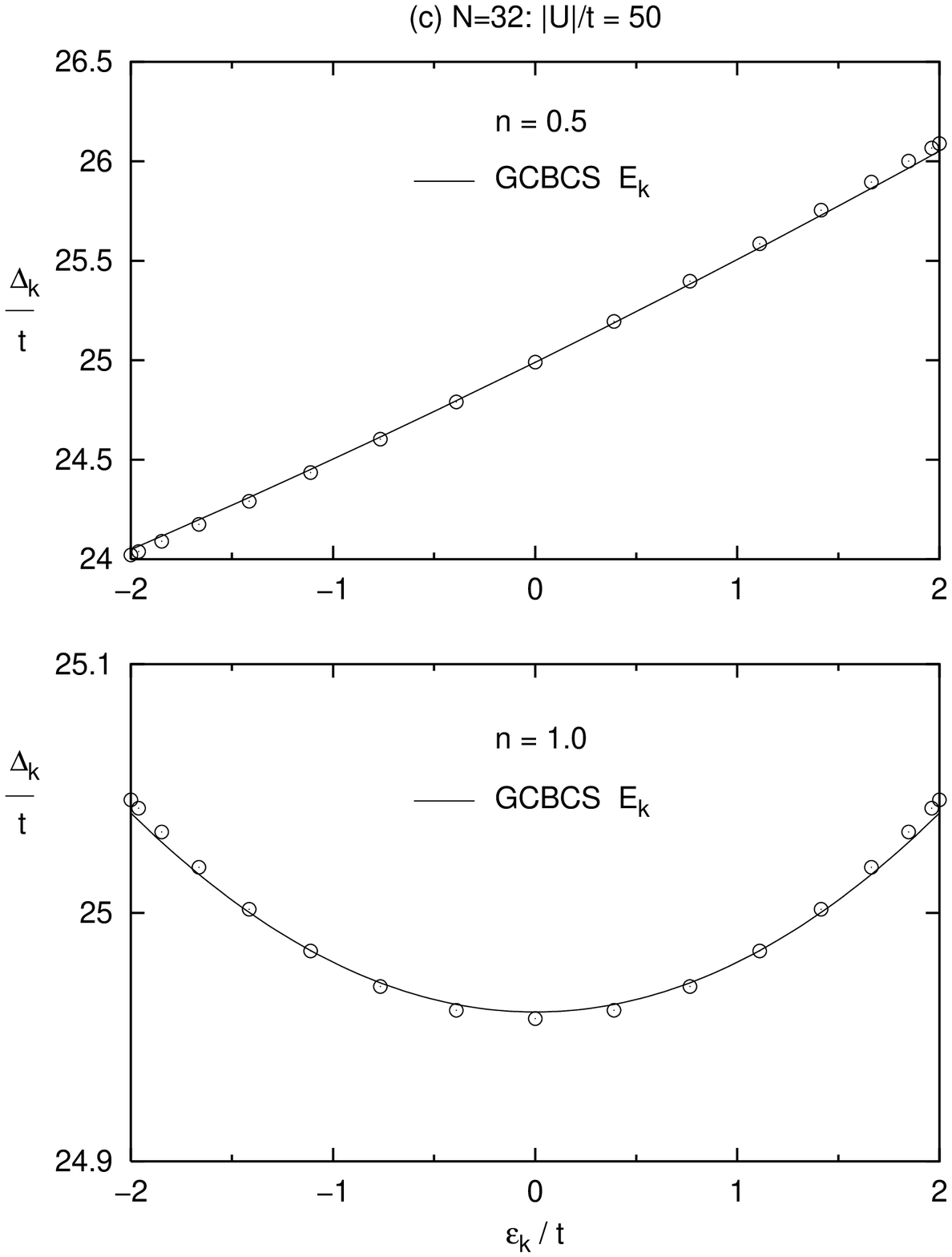}
\caption{(c) Same as (a), but for $|U|/t=50$.}
\label{fig7c}
\end{figure}

For extremely strong coupling such as $|U|/t=50$,
the optimal momenta for $N_e=15$ and 17 are both zero.
In such a case, the ground state energy increases almost linearly
as a function of $\epsilon_k$ from $k R=0$ to $\pi$.
This can be seen in the upper part of \fig{fig7d}(d), where 
$E_{15}$ and $E_{17}$ are the squares (with the left axis)
and the crosses (with the right axis), respectively.
Accordingly, $\Delta_k$ is minimum at $k R=0$ and 
increases linearly as a function of $\epsilon_k$, as seen in \fig{fig7c}(c).
In the grand canonical case, $\tilde \mu$ (related to $\mu$ by \eq{hart})
for such strong coupling is a large negative value, and $E_k$ has a minimum
well below the bottom of the band.

\addtocounter{figure}{-1}
\begin{figure}
\psbb{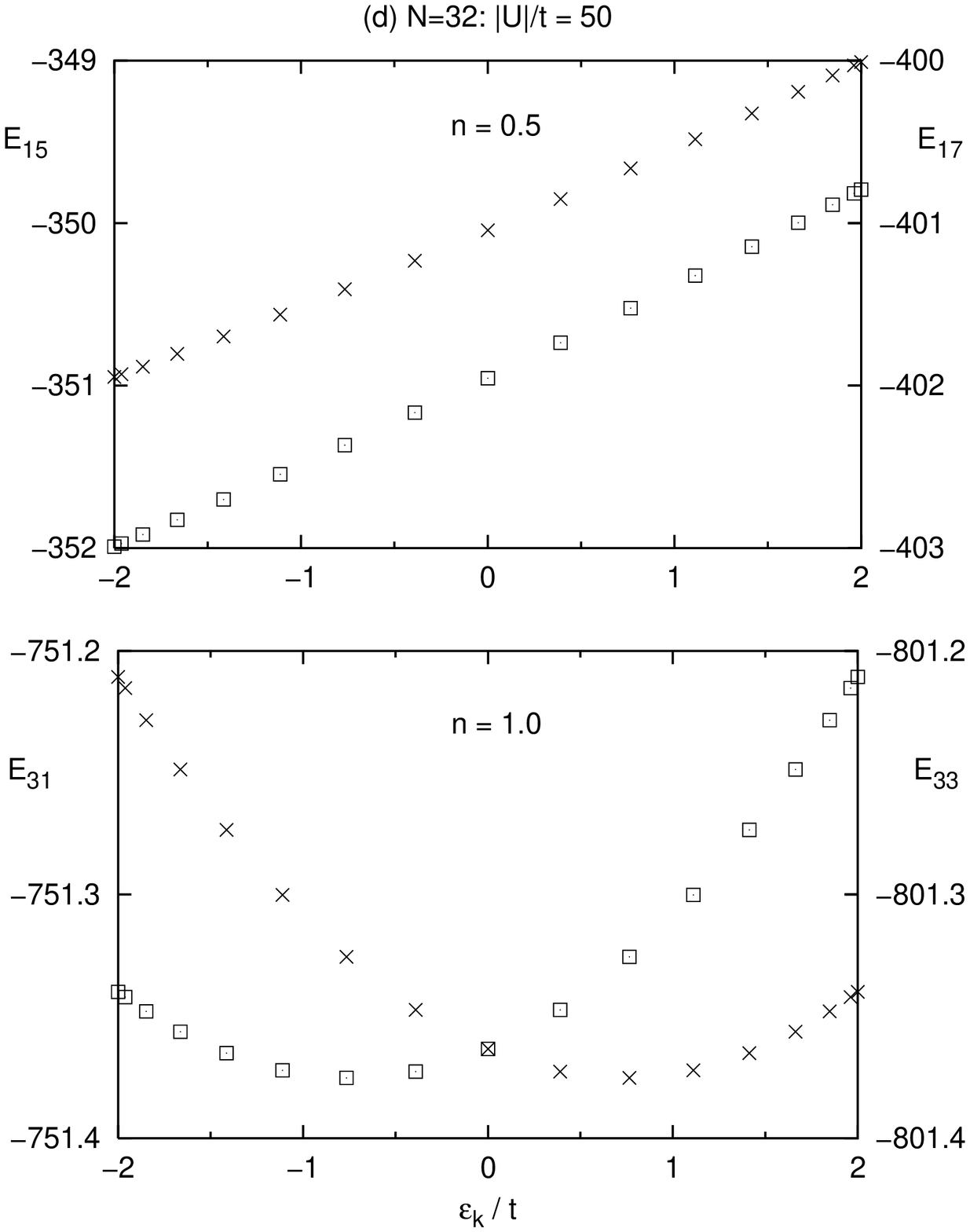}
\caption{(d) Ground state energy for odd $N_e$ as a function of the 
quasi-particle kinetic energy $\epsilon_k$, for $N=32$.
In the upper figure, the energies for $N_e=15$ and 17 are shown with the
squares (with the left axis) and the crosses (with the right axis), respectively; in the lower figure, the energies for $N_e=31$ and 33 are plotted 
accordingly.}
\label{fig7d}
\end{figure}

Half filling is a special case due to the particle-hole symmetry.
In this case $\tilde \mu$ is zero for any coupling strength
and hence the quasi-particle energy is always minimum at $|k R|=\pi/2$.
By the canonical variation,
the ground state energy for $N_e=31$ (the largest odd number for $n\le 1$)
has its minimum at $3\pi/8$.
This can be seen in the lower part of \fig{fig7d}(d) 
(the squares, with the left axis),
although the differences in energy for different $|k|$'s are rather small.
The $E_{33}$ (the crosses, with the right axis) is the mirror image of 
$E_{31}$ about $|k R|=\pi/2$ as a function of the momentum (for any $|U|$), 
as follows from the particle-hole symmetry relation \cite{lieb68}.
Interestingly, the optimal momentum for $N_e=31$ (and for $N_e=33$)
is still close to $\pi/2$ for this coupling strength.
Indeed, one requires even stronger coupling to
have $E_{31}$ that behaves like $E_{15}$ in \fig{fig7d}(d), while
$E_{33}$ will decrease in a symmetric fashion linearly from $k R=0$ to $\pi$.

As the coupling increases, the system of any finite size 
approaches the ``bulk'' limit.
We have seen above for $n=0.5$ that in the strong-coupling limit,
the energies for both of the odd systems (i.e., $N_e=15$ and 17) 
have their minimum values with the momentum $k=0$.
This is true for any finite size and any density smaller than unity.
Thus the canonical $\Delta(N_e)$ with the lowest $E_{N_e+1}$ and $E_{N_e-1}$
($N_e$ even) converges to $\Delta_\circ$, which is, for such strong coupling,
given by $\sqrt{(\epsilon_{\rm min} - \tilde\mu)^2 + \Delta_{\rm BCS}^2}$.
This is not the case for half filling, however:
the lowest $E_{31}$ and $E_{33}$ will have $k R=0$ and $\pi$, respectively,
while both should have $k R=\pi/2$ for the canonical gap to converge to $\Delta_\circ$.

For $4 m + 2$ electrons,
the agreement of $\Delta_k$ with $E_k$ is not as good for all quasi-particle
momenta for weak coupling.
In the strong-coupling limit, however, $\Delta_k$ eventually
converges to $E_k$ for all the momenta as we have seen above for $N_e=4 m$.
Finally, as explained above for the super-even effect,
the minimum gap for $N_e=4 m + 2$ for weak coupling has different $|k|$'s
for the $N_e+1$ and $N_e-1$ systems.  On the contrary, the grand canonical
gap has always the same momentum for an electron and a hole. This explains 
why it does not reproduce the exact gap for $N_e=4 m + 2$ for such weak
coupling, while the canonical one does.

\subsection{Occupation probability}
\label{sec:res:occ}

\begin{figure}
\ps{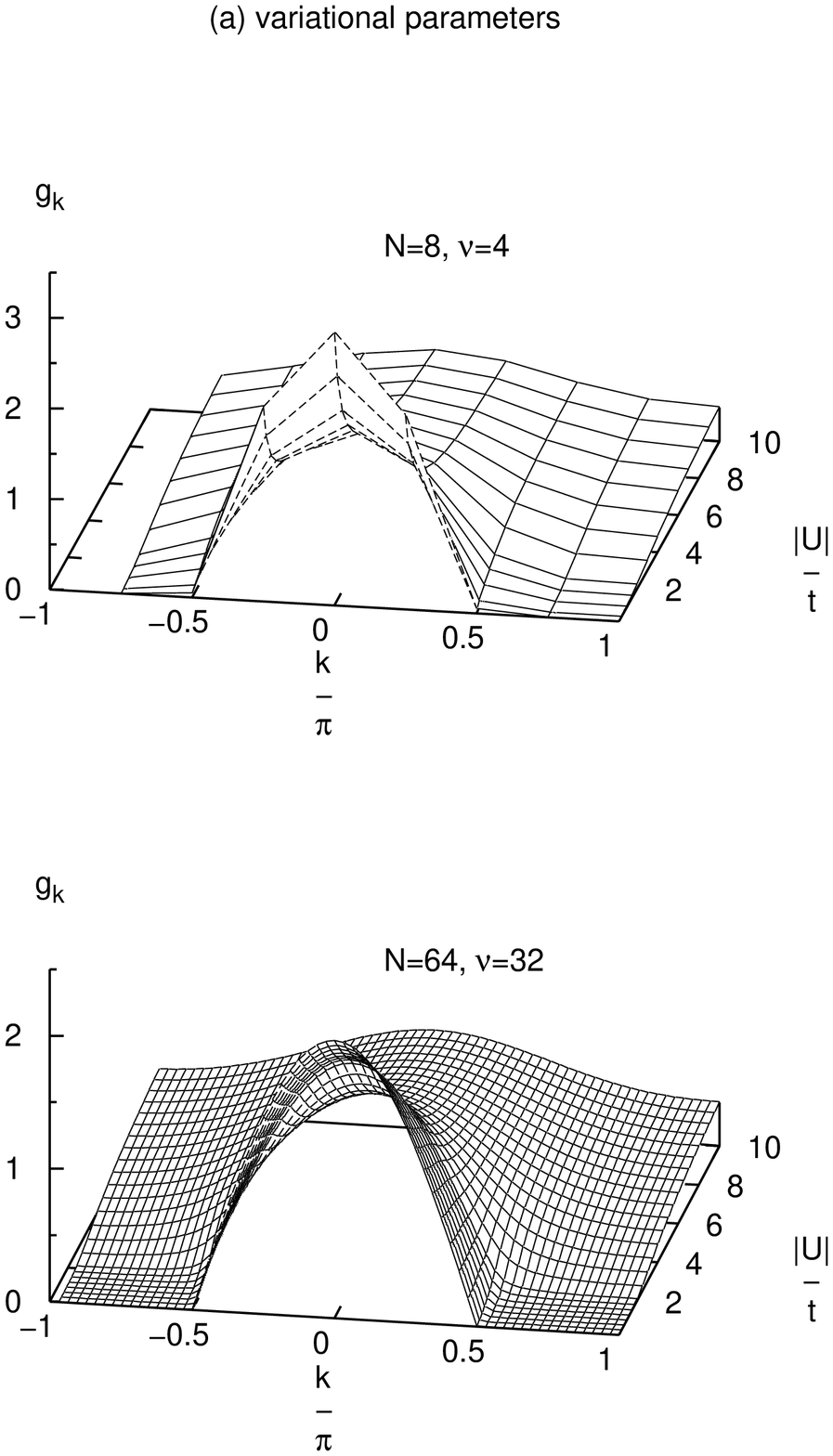}
\caption{(a) Variational parameter $\{g_k\}$ as a function of $k$ and the
coupling strength $|U|/t$, for $N=8$ (upper figure) and 64 (lower figure)
and for half filling.  The increment in $|U|/t$ has been taken to be finer for
$N=64$, while the increment in $k$ is naturally smaller.}
\label{fig8a}
\end{figure}

To conclude the discussion of our results, we show in \fig{fig8a}(a)
the variational parameter $\{g_k\}$ and (b) the occupation probability 
$\{n_k\}$ as a function of $k$ and the coupling strength $|U|$,
for $N=8$ and 64 and for half filling.
In both figures, the points at discrete values of $k$ (and discrete values of $|U|$) are simply connected by lines.
The smallest $|U|$ shown in these figures is $0.25\,t$, and for $N=64$,
the increment in $|U|$ has been taken finer than for $N=8$ (while the
$k$ increment is naturally smaller for larger system size).
The results shown have been obtained by the canonical BCS formalism.
However, also in the grand canonical scheme, 
both $\{g_k\}$ and $\{n_k\}$ have the same overall shape
as a function of $k$ and $|U|$: the actual values of $\{g_k\}$ may be
different but only slightly, so that they look the same as those 
in \fig{fig8a}(a) in the given scale.

We can see in \fig{fig8a}(a) that in the strong-coupling limit, 
$\{g_k\}$ as a function of $k$ is almost flat (in the given scale) 
for a given $|U|$,
and thus all the unperturbed levels are almost equally mixed \cite{nozieres85}.
On the other hand,
in the zero-coupling limit, $\{g_k\}$ becomes a cosine function of $k$.
Since the grand canonical BCS yields the same behaviour of $\{g_k\}$,
this can be understood by the simple expression of \eq{soln}.
As $|U|$ approaches zero, $\Delta_{\rm BCS}$ goes to zero, and 
$E_k$ can be expanded up to the leading order in $\Delta_{\rm BCS}$.
Then \eq{soln} reduces to
\be
g_{\bk} = {1\over\Delta_{\rm BCS}}\,\biggl[\,
|\epsilon_{\bk}-\tilde\mu|\,\biggl(1+{\Delta_{\rm BCS}^2\over 2\,(\epsilon_{\bk}-\tilde\mu)^2}\biggr)-(\epsilon_{\bk} - \tilde\mu)\,\biggr]\;.
\ee
Thus in the limit of $|U|\rightarrow 0$ and $\Delta_{\rm BCS}\rightarrow 0$,
\bea
g_{\bk}&\simeq&{\Delta_{\rm BCS}\over 2\,|\epsilon_{\bk}-\tilde\mu|}\quad
(\epsilon_{\bk}>\tilde\mu)\nonumber\\{\rm and}&&\\
g_{\bk}&\simeq&{2\,|\epsilon_{\bk}-\tilde\mu|\over \Delta_{\rm BCS}}\quad
(\epsilon_{\bk}<\tilde\mu)\;.\nonumber
\eea
Hence $g_{\bk}\rightarrow 0$ for $\epsilon_{\bk}>\tilde\mu$, while
for $\epsilon_{\bk}<\tilde\mu$, $g_{\bk}$ diverges in the zero-coupling limit,
and provides an image of $\epsilon_{\bk}$ (note that for half filling, $\tilde\mu=0$).

Furthermore, the occupation probability clearly shows how 
the distribution over the unperturbed states changes as a function of
the coupling strength.
It can be seen in \fig{fig8b}(b) that
the distribution function for the non-interacting case is recovered 
for weak coupling; $n_k=1$ for $\epsilon_k<\tilde\mu$
and 0 for $\epsilon_k>\tilde\mu$, and $n_k=0.5$ 
at the doubly degenerate Fermi level.
In \fig{fig8b}(b), not only the scale is reduced compared to \fig{fig8a}(a),
but also the relative (average) height of $n_k$ at $|U|/t=10$ against 
the one in the zero-coupling limit is larger.
Thus $n_k$ may appear to have more variation as a function of $k$
in the strong-coupling limit than $g_k$ does.
However, the difference between the maximum and minimum values at $|U|/t=10$
are approximately the same for both cases, and $n_k$ is about 0.7
at $k=0$ and about 0.3 at the edges of the band.
As the coupling is made stronger, the occupation probability (and $g_k$)
becomes almost equal for all the states, and $n_k\simeq 0.5$ for all $k$
in the case of half filling.

\addtocounter{figure}{-1}
\begin{figure}
\ps{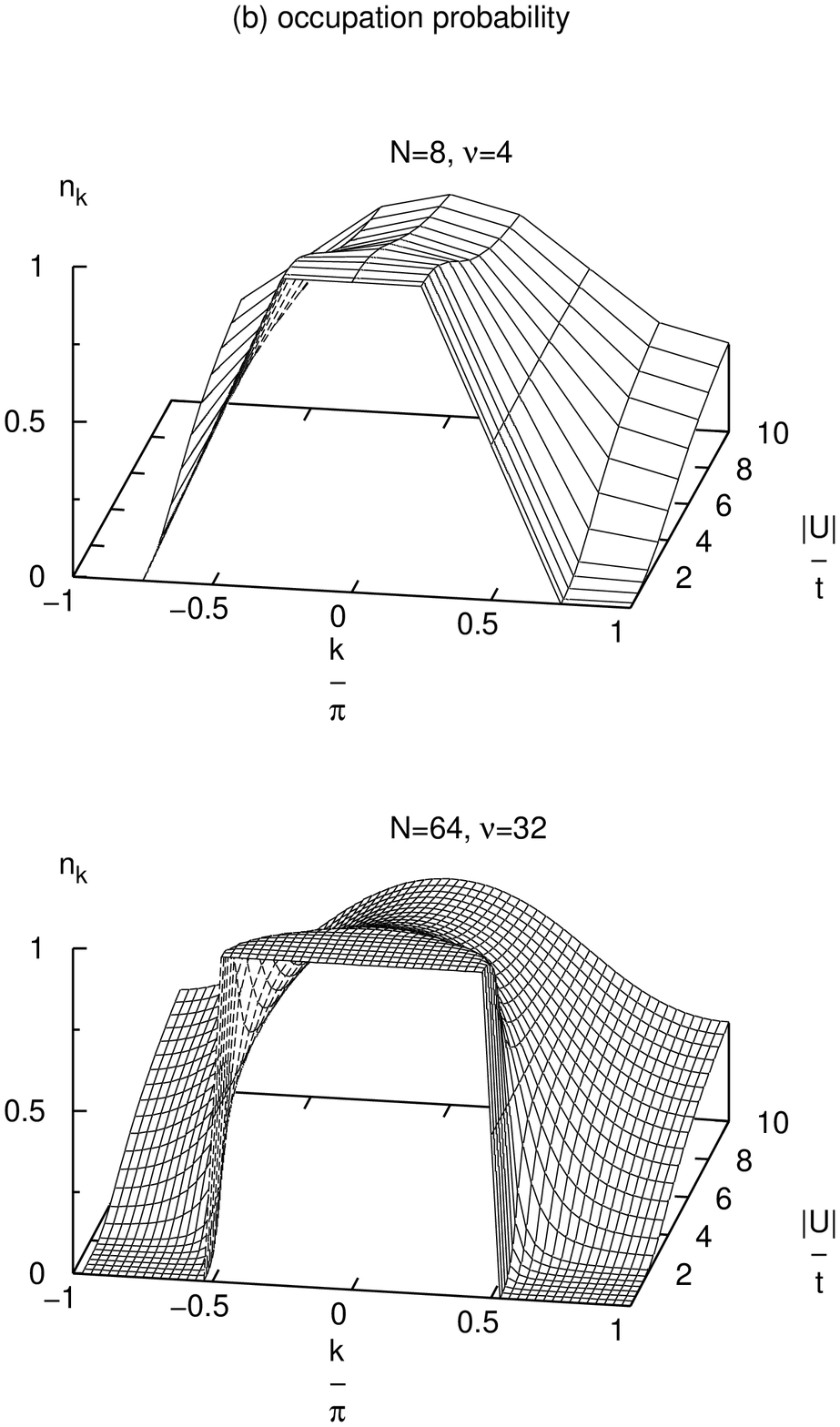}
\caption{(b) Occupation probability $\{n_k\}$ as a function of $k$ and
$|U|/t$, for $N=8$ (upper figure) and 64 (lower figure) and for half filling.
At $|U|/t=10$, $n_k$ is roughly 0.7 and 0.3 at $k=0$ and $\pi$, respectively
(here the lattice constant $R\equiv 1$).
In the weak-coupling limit, the distribution for the non-interacting electron
gas is recovered.}
\label{fig8b}
\end{figure}

\section{SUMMARY AND DISCUSSIONS}
\label{sec:dis}

We have formulated BCS theory for a canonical ensemble, following earlier work
by Dietrich {\it et al.} \cite{dietrich64} and Falicov and Proetto \cite{falicov93},
and very recently by Braun and von Delft  \cite{braun98}.
We have also generalized the linear formulation introduced in Ref. \cite{falicov93} for any system size.  However, this method has proven to be numerically
too intensive for ``large'' systems, and provided at best only marginal improvement over
the nonlinear canonical formulation. \par

In this work we have adopted a very definite model, the attractive Hubbard model, for
various reasons. First, we wanted to have an exact solution with which to monitor
the improvement over the grand canonical scheme. These are available for the Hubbard
model in one dimension by the Bethe Ansatz technique, and so we have used these as
a benchmark throughout this work. Second, we wanted to study a system which, by
choice of parameters, could easily span the weak coupling to strong coupling regime,
as well as the low density to high density limits.
In this way we have observed the crossover from the bulk to quantum limit for a variety of
regimes.
The attractive Hubbard model
is no doubt the ``minimal'' model that accomplishes this. Finally, we wanted to
use a model which could readily be generalized to a realistic case, so, for example,
one could use a parametrized tight-binding model to fit the band structure for
Al to better describe ultrasmall metallic Al grains.\par

In this work, however, we focused on the first two points listed above, with the
goal of establishing generic trends. The existence of the parity effect, 
as already determined by a parity-projected grand canonical scheme 
\cite{janko94,braun97} and more
recently by the canonical formulation of a uniformly-spaced-level model \cite{braun98},
emerges quite naturally in this model.  
However, in addition to the even-odd effect,
we have found a {\it super-even} effect, with oscillations in the superconducting
gap occurring between even electron numbers which are multiples of 4 ($4m$) and
non-multiples of 4 ($4m + 2$). The magnitude of the gap variation for even 
numbers of electrons is, in some cases, comparable to the even-odd variation, i.e.
$|\Delta(4m) - \Delta(4m+1)| \approx |\Delta(4m) - \Delta(4m+2)|$. 
Thus, such oscillations should be observable in the same kind of tunneling 
experiments for seeing the even-odd effect. 
We note, however, a key ingredient 
for the super-even effect to occur 
is the double degeneracy of levels, which in our case comes about from the
simple equality, $\epsilon_{\bf k} = \epsilon_{\bf -k}$. In a more general case,
the degeneracy structure will be more complicated, and therefore oscillations will
exist but may not be as simply periodic as a function of electron number as is the
case here.
The super-even effect is also a result of quantized energy levels due to 
finite system size, and the effect will be stronger for smaller systems.
For very weak coupling, the grand canonical BCS fails to reproduce the 
super-even effect, while the canonical scheme does; indeed, it yields 
very good agreement with the exact solutions.
\par

Finally, we note that the grand canonical BCS quasi-particle dispersion relation was
beautifully reproduced by the canonical results, {\it simply by varying the odd
number ground state momentum}, with the proviso that the electron and hole
momenta were the same. This simple correspondence is quite surprising.\par

\acknowledgments

This research was supported by the Avadh Bhatia Fellowship and by the
Natural Sciences and Engineering Research Council of Canada and
the Canadian Institute for Advanced Research.
We are grateful to Bob Teshima for his help in efficiently evaluating the residue integrals using the analytical results.
K.T. thanks Rajat Bhaduri, Jules Carbotte and Elisabeth Nicol for helpful 
comments and discussions.
Calculations were performed on the 42 node SGI
parallel processor at the University of Alberta.

\appendix
\section*{Linear Canonical Variation}
\label{appendix}

We summarize the linearized formulation of the canonical variation.
In addition to the wave function for $N_e=2\nu$ defined by \eq{cwfe}, 
we define the wave function for odd number of electrons $N_e=2\nu+1$, 
following \cite{marsiglio97} as
\bea
|\Psi_{2\nu+1}\rangle=&a_{{\bf q}\sigma}^\dagger&
\sum_{{\bf k}_1\neq{\bf q}\,<}\sum_{{\bf k}_2\neq{\bf q}\,<}
\cdots\sum_{<\,{\bf k}_\nu\neq{\bf q}}\,
C({\bf k}_1,{\bf k}_2,\cdots,{\bf k}_\nu)\nonumber\\
&\times&\prod_{i=1}^{\nu}\,a_{{\bf k}_i\uparrow}^\dagger
a_{-{\bf k}_i\downarrow}^\dagger\,|0\rangle\;,\label{cwfo}
\eea
where spin of the extra electron $\sigma$ can be either up or down.
The extra electron blocks the state $\bf q$ from being occupied by
pairs.  There is no variational parameter for the blocked state:
for the variational calculations, we always choose $\bf q$ that
gives the lowest energy.
Note that the $C$'s in the above equation are different from those
for the same number of pairs in \eq{cwfe} due to the missing $\bf q$.

We now derive the variational equation using
$|\Psi_\nu\rangle$ in \eq{cwfe} for an even number of electrons $N_e=2\nu$.
The formulae below are slightly modified
for an odd number of electrons $N_e=2\nu+1$
in terms of the odd wave function \eqn{cwfo}.  We mention the difference
for $N_e=2\nu+1$ in the end.

The normalization factor for $|\Psi_{2\nu}\rangle$ defined in \eq{cwfe} is
\bea
\langle\Psi_{2\nu}|\Psi_{2\nu}\rangle&=&
\sum_{{\bf k}_1\,<}\cdots\sum_{<\,{\bf k}_\nu}\;
\sum_{{\bf p}_1\,<}\cdots\sum_{<\,{\bf p}_\nu}\,\nonumber\\
&&C^*({\bf k}_1,\cdots,{\bf k}_\nu)\;
C({\bf p}_1,\cdots,{\bf p}_\nu)\nonumber\\
&\times&\langle 0|\,a_{-{\bf k}_\nu\downarrow} a_{{\bf k}_\nu\uparrow}\,
\cdots\,a_{-{\bf k}_1\downarrow} a_{{\bf k}_1\uparrow}\,\nonumber\\
&&\hspace{.5cm}a_{{\bf p}_1\uparrow}^\dagger a_{-{\bf p}_1\downarrow}^\dagger\,\cdots\,
a_{{\bf p}_\nu\uparrow}^\dagger a_{-{\bf p}_\nu\downarrow}^\dagger\,|0\rangle
\nonumber\\\nonumber\\
&=&\sum_{{\bf k}_1\,<}\cdots\sum_{<\,{\bf k}_\nu}\,
|\,C({\bf k}_1,\cdots,{\bf k}_\nu)\,|^2\;.
\label{cforme_norm}
\eea
Similarly, the expectation value of the kinetic energy operator
in $H=\hat T+\hat V$ is
\bea
&&\langle\Psi_{2\nu}|\,\hat{T}\,|\Psi_{2\nu}\rangle=\langle\Psi_{2\nu}|\,
\sum_{\bf{k}\sigma} \epsilon_{\bf{k}}\,a_{\bf{k}\sigma}^\dagger
a_{\bf{k}\sigma}\,|\Psi_{2\nu}\rangle\nonumber\\
&=&\sum_{{\bf k}_1\,<}\cdots\sum_{<\,{\bf k}_\nu}\,
|\,C({\bf k}_1,\cdots,{\bf k}_\nu)\,|^2\;2\,
(\epsilon_{\bf{k}_1}+\cdots+\epsilon_{\bf{k}_\nu})\;.
\label{cforme_ke}
\eea
Calculation of the potential energy term is more tedious. After some
lengthy operator algebra it can be written in a general form for
$\nu$ pairs as
\bea
&&\langle\Psi_{2\nu}|\,\hat{V}\,|\Psi_{2\nu}\rangle\nonumber\\\nonumber\\&=&
- {|U| \over N}\,\langle\Psi_{2\nu}|\,\sum_{\bf{k}\bf{k}'\bf{l}}
\,a_{{\bf k}\uparrow}^\dagger a_{-{\bf k} + {\bf l} \downarrow}^\dagger 
a_{-{\bf k}' + {\bf l}\downarrow}a_{{\bf k}'\uparrow}
\,|\Psi_{2\nu}\rangle\nonumber\\\nonumber\\
&=&-{|U|\over N}\,\sum_{{\bf k}_1\,<}\cdots\sum_{<\,{\bf k}_\nu}\,
C({\bf k}_1,\cdots,{\bf k}_\nu)\,\nonumber\\
&\times&\Bigl[\;\nu\,(\nu-1)\,
C^*({\bf k}_1,\cdots,{\bf k}_\nu)\nonumber\\&+&\sum_{{\bf p}<{\bf k}_1}
C^*({\bf p},{\bf k}_1,{\bf k}_2,\cdots,{\bf k}_{\nu-1})\,\nonumber\\
&+&\sum_{{\bf k}_1<{\bf p}<{\bf k}_2}
C^*({\bf k}_1,{\bf p},{\bf k}_2,\cdots,{\bf k}_{\nu-1})\nonumber\\
&+&\quad\cdots\quad+\sum_{{\bf k}_{\nu-1}<{\bf p}}
C^*({\bf k}_1,{\bf k}_2,\cdots,{\bf k}_{\nu-1},{\bf p})\nonumber\\
&+&\sum_{{\bf p}<{\bf k}_1}C^*({\bf p},{\bf k}_1,{\bf k}_2,\cdots,
{\bf k}_{\nu-2},{\bf k}_\nu)\,\nonumber\\
&+&\sum_{{\bf k}_1<{\bf p}<{\bf k}_2}C^*({\bf k}_1,{\bf p},{\bf k}_2,\cdots,
{\bf k}_{\nu-2},{\bf k}_\nu)\nonumber\\
&+&\quad\cdots\quad+\sum_{{\bf k}_{\nu}<{\bf p}}C^*({\bf k}_1,{\bf k}_2,\cdots,
{\bf k}_{\nu-2},{\bf k}_\nu,{\bf p})\nonumber\\
&+&\quad\cdots\nonumber\\
&+&\sum_{{\bf p}<{\bf k}_2}C^*({\bf p},{\bf k}_2,{\bf k}_3,\cdots,{\bf k}_\nu)\,\nonumber\\
&+&\sum_{{\bf k}_2<{\bf p}<{\bf k}_3}C^*({\bf k}_2,{\bf p},{\bf k}_3,\cdots,
{\bf k}_\nu)\nonumber\\
&+&\quad\cdots\quad+\sum_{{\bf k}_\nu<{\bf p}}C^*({\bf k}_2,{\bf k}_3,\cdots,
{\bf k}_\nu,{\bf p})\;\Bigr]
\label{cforme_pe}
\eea
The first term on the right hand
side corresponds to a Hartree-like term, and gives the interaction energy of
each of the $\nu$ pairs with the other $\nu - 1$ pairs. The other terms
systematically consider the various cases when one pair is scattered into
another (unoccupied) pair state.

In the following, we take the $C$'s to be real variables.
In \refer{dietrich64} it has been proved for the BCS formulation
that for negative pairing-type interactions,
real (and positive) variational parameters yield the lowest energy.
The minimization condition for the energy \eqn{venergy}
with respect to the real $\{C\}$ results in the equation
\bea
C({\bf k}_1,\cdots,{\bf k}_\nu)&=&-{1\over d}\,{|U|\over N}\,\Bigl[
\sum_{{\bf p}<{\bf k}_1}
C({\bf p},{\bf k}_1,{\bf k}_2,\cdots,{\bf k}_{\nu-1})\,\nonumber\\
&+&\sum_{{\bf k}_1<{\bf p}<{\bf k}_2}
C({\bf k}_1,{\bf p},{\bf k}_2,\cdots,{\bf k}_{\nu-1})\nonumber\\
+\quad\cdots\quad&+&\sum_{{\bf k}_{\nu-1}<{\bf p}\neq{\bf k}_\nu}
C({\bf k}_1,{\bf k}_2,\cdots,{\bf k}_{\nu-1},{\bf p})\nonumber\\
&+&\sum_{{\bf p}<{\bf k}_1}C({\bf p},{\bf k}_1,{\bf k}_2,\cdots,
{\bf k}_{\nu-2},{\bf k}_\nu)\,\nonumber\\
&+&\sum_{{\bf k}_1<{\bf p}<{\bf k}_2}C({\bf k}_1,{\bf p},{\bf k}_2,\cdots,
{\bf k}_{\nu-2},{\bf k}_\nu)\nonumber\\+\quad\cdots\quad
&+&\sum_{{\bf k}_{\nu-2}<{\bf p}\neq{\bf k}_{\nu-1}<{\bf k}_\nu}
C({\bf k}_1,{\bf k}_2,\cdots,{\bf k}_{\nu-2},{\bf p},{\bf k}_\nu)\nonumber\\
&+&\sum_{{\bf k}_{\nu}<{\bf p}}C({\bf k}_1,{\bf k}_2,\cdots,
{\bf k}_{\nu-2},{\bf k}_\nu,{\bf p})\nonumber\\
&+&\cdots\nonumber\\&+&\sum_{{\bf p}\neq{\bf k}_1<{\bf k}_2}
C({\bf p},{\bf k}_2,{\bf k}_3,\cdots,{\bf k}_\nu)\,\nonumber\\
&+&\sum_{{\bf k}_2<{\bf p}<{\bf k}_3}C({\bf k}_2,{\bf p},{\bf k}_3,\cdots,
{\bf k}_\nu)\nonumber\\
+\quad\cdots\quad&+&\sum_{{\bf k}_\nu<{\bf p}}C({\bf k}_2,{\bf k}_3,\cdots,
{\bf k}_\nu,{\bf p})\;\Bigr]\;,\label{csol}
\eea 
where
\bea
d=E_\nu-2\,(\,\epsilon_{{\bf k}_1}+\cdots+\epsilon_{{\bf k}_\nu}\,)
+{|U|\over N}\,\nu^2\;.
\label{cdenom}\eqnum{12$^\prime$}
\eea
This should be solved for all the $C$'s selfconsistently.

For an odd number of electrons,
in the energy expectation value and the variational equation
in terms of $|\Psi_{2\nu+1}\rangle$, the blocked $\bf q$ is excluded in all
the $\{{\bf k}\}$ sums and $\{{\bf p}\}$ sums.
Other than that, the normalization factor for $|\Psi_{2\nu+1}\rangle$ is
the same as \eq{cforme_norm}: the kinetic energy term is the same
as \eq{cforme_ke} except that
$2 (\epsilon_{\bf{k}_1}+\cdots+\epsilon_{\bf{k}_\nu})$ should be replaced by
$2 (\epsilon_{\bf{k}_1}+\cdots+\epsilon_{\bf{k}_\nu})+\epsilon_{\bf q}$:
the potential energy term is the same as \eq{cforme_pe} except that
the factor $\nu\,(\nu-1)$ should be replaced by $\nu^2$: the variational
equation is the same as \eq{csol}, while in $d$ as defined in \eq{cdenom}
$-\,\epsilon_{\bf q}$ should be added and $\nu^2$ should be replaced by
$\nu\,(\nu+1)$.  

\begin{figure}
\psbb{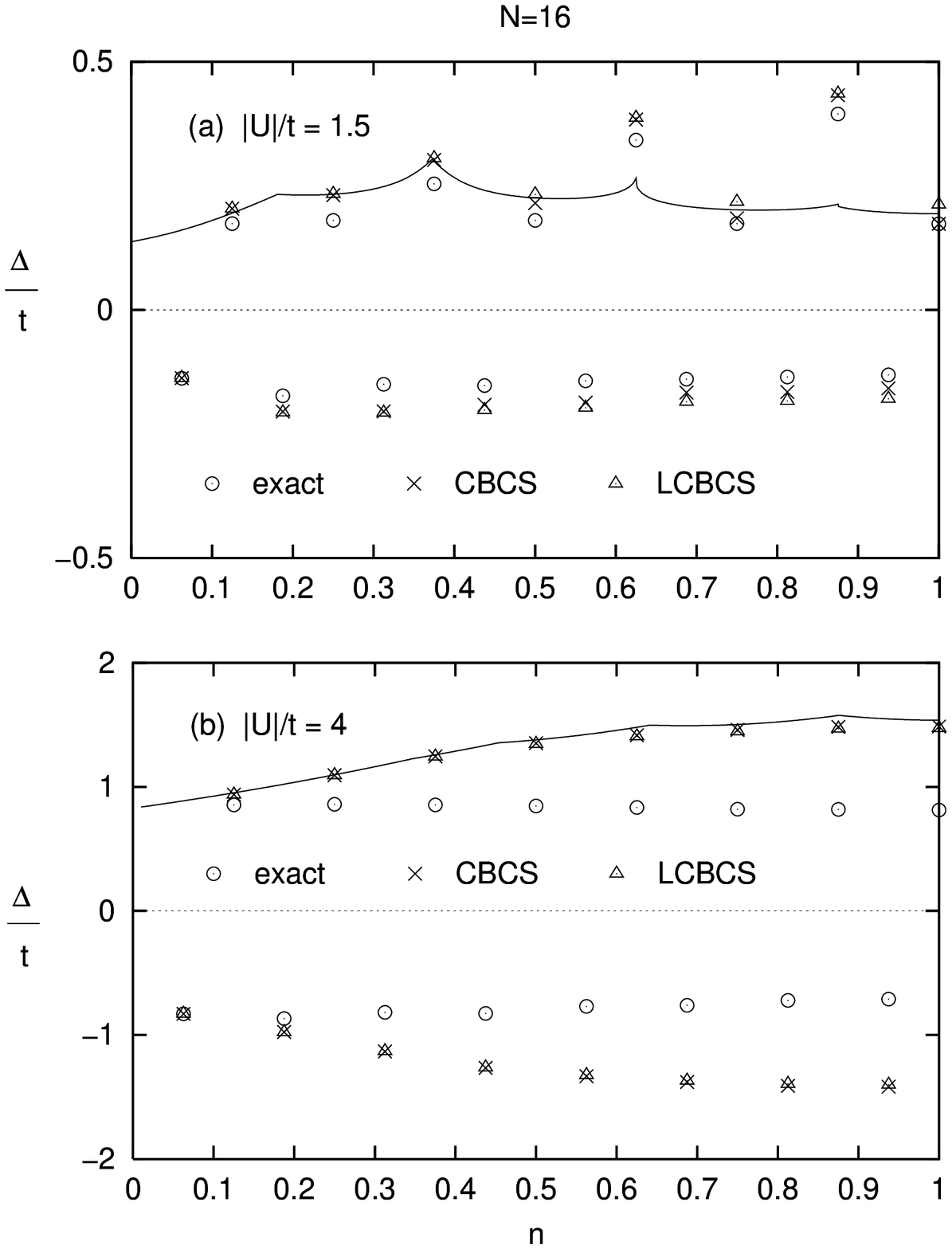}
\caption{Energy gap as a function of the electron density $n$ for $N=16$, for
(a) $|U|/t=1.5$ and (b) $|U|/t=4$.  The exact solutions are shown with circles,
while the canonical results from the $g$-formulation and from the linear
$C$-formulation are shown with the crosses and triangles, respectively.}
\label{fig9}
\end{figure}

We have found that this linear variation in terms of the $C$'s just barely 
improves the ground state energy, compared to the nonlinear variation in 
terms of the $g$'s.  Interestingly, the $C$-formulation does not 
improve the $g$-formulation for the energy gap.  We illustrate this in
\fig{fig9} for $N=16$ and for $|U|/t=1.5$ and 4.
We found that for smaller system size, the two formulations do not make any
difference for any coupling strength, for the gap as well as the ground state
energy.  For larger system size
(which was limited for $N\alt 30$) the gap from the $C$-formulation is
slightly worse than the one from the $g$-formulation for weak coupling
and larger density.
This can be seen in \fig{fig9}(a), while for $|U|/t=4$ in (b) the two results
have converged for all the density.

\end{document}